\catcode`\@=11  
\def\answ{o}    
\def\onecol{o }      
\def\twocol{t }      
\def\prepri{p }      
\def\nofig{n }       

\documentstyle[aps,eqsecnum,floats,epsf]{revtex}
\begin{document}
\draft
\if\answ\twocol
      \twocolumn[\hsize\textwidth\columnwidth\hsize\csname
             @twocolumnfalse\endcsname
\fi

\title{Hyperbolic formulations and numerical relativity: \\
Experiments using Ashtekar's connection variables}
\author{Hisa-aki Shinkai  \cite{Email-HS}}
\address{
Centre for Gravitational Physics and Geometry,
104 Davey Lab., Department of Physics,\\
The Pennsylvania State University,
University Park, Pennsylvania 16802-6300, USA
}
\author{Gen Yoneda  \cite{Email-GY}}
\address{
Department of Mathematical Sciences, Waseda University,
Shinjuku, Tokyo,  169-8555, Japan
}
\date{September 6, 2000}
\maketitle

\begin{abstract}
\if\answ\twocol
     \widetext
\fi
In order to perform accurate and stable long-time numerical integration
of the Einstein equation, several hyperbolic systems have been proposed.
We here present numerical comparisons
between weakly hyperbolic, strongly hyperbolic,
and symmetric hyperbolic systems based on Ashtekar's connection variables.
The primary advantage for using this connection formulation in this experiment
is that we can keep using the same dynamical variables for all levels of
hyperbolicity.
Our numerical code demonstrates gravitational wave propagation in
plane symmetric spacetimes, and we compare the
accuracy of the simulation by monitoring
the violation of the constraints.
By comparing with results obtained from the weakly hyperbolic system, we
observe the strongly and symmetric hyperbolic system show better
numerical performance (yield less constraint violation),
but not so much difference between the latter two.
Rather, we find that the
symmetric hyperbolic system is not always the best in numerical performances.

This study is the premier to present full numerical simulations
using Ashtekar's variables.  We also describe our procedures in detail.

\end{abstract}

\pacs{gr-qc/0005003, revised version}
\if\answ\twocol
     \vskip 2pc]
     \narrowtext
\fi
\baselineskip = 13pt
\section{Introduction}

Numerical relativity -- solving the Einstein equation numerically --
is now an essential field in gravity research.
As is well known, critical collapse in gravity systems was
first discovered by numerical simulation \cite{choptuik}.
The current mainstream of numerical relativity is to
demonstrate the final phase of compact
binary objects
related to gravitational wave observations \cite{PTPsupple}, and
these efforts are
now again shedding light on the
mathematical structure of the Einstein equations.


Up to a couple of years ago, the standard Arnowitt-Deser-Misner (ADM)
decomposition of the Einstein
equation was taken as the standard formulation for numerical relativists.
Difficulties in accurate/stable long-term
evolutions were supposed to be overcome by choosing proper gauge
conditions and boundary conditions.  Recently, however, several
numerical experiments show
that the standard ADM is not the best formulation for numerics, and
finding a better formulation
has become one of the main research topics.
\footnote{
Note that we are only concerned with the free evolution system of the
initial data, that is, we only solve the
constraint equations on the initial hypersurface.
The accuracy and/or stability of the system is normally observed by
monitoring the violation of constraints during the free evolution.}

One direction in the community is to apply
conformally decoupled and tracefree re-formulation of ADM
system which were
first used by Nakamura {\it et al.}  \cite{SN}.
The usefulness of this re-formulation were
confirmed by another groups to show a long-term
stable numerical evolutions.
\cite{BS,AEI00}.
Although there is an effort to show why this
re-formulation is better than ADM
\cite{AABSS},
we do not yet know this method is robust for all situations.

Another alternative approach to ADM is to formulate the
Einstein equations to reveal
hyperbolicity \cite{reviewhyp}.
A certain kind of hyperbolicity of the dynamical
equations is essential
to analyze their propagation features mathematically,
and are known to
be useful in numerical approximations (we explain
these points in \S 2).
The propagation of the original ADM constraint equations obeys
well-posed behavior \cite{Fri-con}
but the dynamical equations of the ADM system are not a hyperbolic
system at all
(and these facts can be also applied to the conformally
decoupled version).
Several hyperbolic formulations have been proposed to
re-express the Einstein
equation, with different levels:
weakly, strongly and symmetric hyperbolic systems
(we will discuss in detail in \S 2).
Several numerical tests were also performed in this
direction, and we can see advantages in numerical stability over the
original ADM system (e.g.
tests \cite{BMSS95} of Bona-Mass\'o's symmetrizable
form \cite{BM92}, tests \cite{SBCSThyper} of Choquet-Bruhat and
York (95)'s symmetrizable form \cite{CBY}), but
the appearance of coordinate shocks is also reported
(\cite{Alcubierre} in the system of \cite{BMSS95}).
A symmetric hyperbolic system of \cite{HF}, on the other hand,
is numerically studied in the context of
``conformal Einstein" approach \cite{conformalEinstein}.

The following questions, therefore, naturally present themselves
(cf. \cite{Stewart}):
(1)
Does hyperbolicity actually contribute to the numerical accuracy/stability?
(2) If so,
which level of hyperbolic formulation is practically useful for numerical
applications? (or does the symmetric hyperbolicity solve all the
difficulties?)
(3) Are there any other approaches to improve the accuracy/stability
   of the system?


In this paper, we try to answer these questions with our
simple numerical experiments.
Such comparisons are appropriate when the fundamental equations
are cast in the same interface, and that is possible at this moment
only using Ashtekar's connection variables \cite{Ashtekar,AshtekarBook}.
More precisely, the authors' recent studies showed the following:
(a) the original set of dynamical equations proposed
by Ashtekar already forms
a weakly hyperbolic system \cite{YS-IJMPD},
(b) by requiring additional gauge conditions {\it or} adding constraints to
the dynamical equations, we can obtain a
strongly hyperbolic system \cite{YS-IJMPD},
(c) by requiring additional gauge conditions {\it and} adding constraints to
the dynamical equations,  we can obtain a symmetric hyperbolic system
\cite{YS-IJMPD,YShypPRL}, and finally
(d) based on the above symmetric hyperbolic system,
we can construct a set of
dynamical systems which is robust against perturbative
errors for constraints and
reality conditions \cite{SY-asympAsh}
({\it aka.} $\lambda$-system \cite{BFHR}).

Based on the above results (a)-(c),
we developed a numerical code which handles
gravitational wave propagation in the
plane symmetric spacetime.
We performed the time evolutions using
   the above three levels of Ashtekar's dynamical equations
   together with the standard ADM equation.
We compare these for accuracy and stability
by monitoring the violation of the constraints.
We also show the demonstrations of our $\lambda$-system (above (d)) in the
succeeding paper (Paper II) \cite{Paper2}, together with new proposal for
controlling the stability.

It is worth remarking that this study is the first one which shows
full numerical simulations of Lorentzian spacetime using
Ashtekar's connection variables.
This research direction was suggested
\cite{AshtekarRomano89} soon after Ashtekar completed
his formulation, but has not yet been completed.
Historically, an application to numerical relativity of the connection
formulation was also suggested \cite{AshtekarBook,Salisbury}
using Capovilla-Dell-Jacobson's version of the connection variables
\cite{CDJ}, which produce an direct relation to Newman-Penrose's $\Psi$s.
However here we apply Ashtekar's original formulation, because we know
how to treat its reality conditions in detail \cite{AshtekarRomanoTate,ys-con},
and how they form hyperbolicities.
We will also describe the basic numerical procedures in this paper.

The outline of this paper is as follows: In the next section, we review
the mathematical background of the hyperbolic formulation briefly and
present our fundamental dynamical equations.
In \S \ref{sec3}, we describe our numerical procedures.
Our experiments are presented in \S \ref{Results1} and \S \ref{Results2},
and we summarize them in \S \ref{Summary}.  The Appendix \ref{appA} is for
showing Ashtekar's basic equations in our notation, and we also present
our experiments based on the Maxwell equation in the Appendix \ref{appB}.
We also introduce briefly
the discussion in our Paper II in the Appendix \ref{appC}.

\section{Hyperbolic Formulations}
\subsection{definitions, properties, mathematical backgrounds}
We say that the system is {\it a first-order (quasi-linear)
partial differential equation system},
if a certain set of
(complex-valued) variables $u_\alpha$ $(\alpha=1,\cdots, n)$
forms
\begin{equation}
\partial_t u_\alpha
= {\cal M}^{l}{}^{\beta}{}_\alpha (u) \, \partial_{l} u_\beta
+{\cal N}_\alpha(u),
\label{def}
\end{equation}
where ${\cal M}$ (the characteristic matrix) and
${\cal N}$ are functions of $u$
but do not include any derivatives of $u$.
If the characteristic matrix is a Hermitian matrix, then we say
(\ref{def}) is {\it a symmetric hyperbolic system}.



Writing the system in a hyperbolic form is the essential
step in proving the system is well-posed.  Here, {\it well-posed}ness
of the system means
($1^\circ$) existence (of at least one solution $u$),
($2^\circ$) uniqueness (i.e., at most solutions), and
($3^\circ$) stability (or continuous dependence of
solutions $\{ u \}$ on the Cauchy data).
The Cauchy problem under weak hyperbolicity is not,
   in general, $C^\infty$ well-posed.
The symmetric hyperbolic system gives us the energy
integral inequalities
which are the primary tools for studying the stability
of the system.
Well-posedness of the symmetric hyperbolic is guaranteed
if the characteristic matrix is independent of $u$,
while if it depends on $u$ we have only the limited proofs for the
well-posedness.
{}From the mathematical point of view,
proving well-posedness with less strict conditions is an
old but active research problem.

We can define another hyperbolic system between
the weakly and symmetric levels.  For example, we say we have a
{\it strongly hyperbolic} (or diagonalizable hyperbolic \cite{YS-IJMPD})
system,
if the characteristic matrix is
diagonalizable and has all real eigenvalues.
The inclusion relation is, then,
\begin{equation}
\mbox{\rm symmetric~hyperbolic} \in
\mbox{\rm strongly~hyperbolic} \in
\mbox{\rm weakly~hyperbolic},
\end{equation}
(which means the symmetric hyperbolicity requires
stronger conditions to be satisfied than the others).
We do not repeat each level's features here (see
\S 2 of \cite{YS-IJMPD}).
However, at the strongly hyperbolic level,
we can prove the finiteness of the energy norm
if the characteristic matrix is independent of $u$
(cf \cite{Stewart}), that is one step definitely advanced than
a weakly hyperbolic form.

{}From the point of numerical applications, to write down the
fundamental equation in an explicitly hyperbolic form
is quite attractive,
not only for its mathematically well-posed features.
It is well known that a certain flux conservative hyperbolic
systems of equations is taken as an essential formulation in the
computational Newtonian hydrodynamics \cite{Hirsch}.
There is also an effort for implementing the boundary condition
by using the characteristic speed (eigenvalues) of the system
\cite{cactus1}.

\subsection{Hyperbolic formulations of the Einstein equation}
As was discussed by Geroch \cite{Geroch},
most physical systems
can be expressed as symmetric hyperbolic systems.
However, the standard ADM system does not form a first
order hyperbolic system.
This can be seen immediately from the fact that the
ADM dynamical equations,
\begin{eqnarray}
{\partial_t} \gamma_{ij} &=& -2 N
K_{ij}+\nabla_jN_{i}+\nabla_i N_{j},
   \label{hatten1}  \\
{\partial_t} K_{ij} &=& N
(~^{(3)\!}R_{ij}+{\rm tr}KK_{ij})-2 N K_{im}K^m_{~j}
-\nabla_i\nabla_j N
\nonumber \\ &~&
+(\nabla_j N^m)K_{mi}+(\nabla_i N^m)K_{mj}+N^m \nabla_m K_{ij},
\label{hatten2}
\end{eqnarray}
have Ricci curvature $~^{(3)\!}R_{ij}$ which
involves second derivatives of
the three-metric  $\gamma_{ij}$ by definition.  (The notation here
is the standard one.
$K_{ij}$ is the extrinsic curvature,
$N$ and $N^i$ are the lapse and shift vector,
respectively. $\nabla$ denotes a covariant derivative on the three-surface.)
For our later convenience, we also write down the
ADM constraint equations,
\begin{eqnarray}
{\cal C}^{\rm ADM}_{H} &:=&
~^{(3)\!}R+(\mbox{tr}K)^2-K_{ij}K^{ij} \approx 0,
\label{conhamilt} \\
{\cal C}^{{\rm ADM} i}_{M} &:=&
\nabla_j(K^{ij}-\gamma^{ij}\mbox{tr}K) \approx 0,
\label{conmoment}
\end{eqnarray}
which are called the Hamiltonian and momentum
constraint equations, respectively.

So far, several first order hyperbolic systems of the Einstein equation
have been proposed;
some of them are symmetrizable (strongly hyperbolic) \cite{BMSS95,BM92}
or symmetric hyperbolic systems
\cite{CBY,HF,FischerMarsden72,FR96}.
There are many variations in the methods for constructing higher
hyperbolic systems, but the number of fundamental
dynamical variables is always results in larger than that of ADM (see a brief
example by Anderson-York(99) in \cite{CBY}).
Several numerical tests
are reported (as we referred in the Introduction) using a particular
hyperbolic formulation, but no numerical
comparisons between these formulations are reported
\cite{hyp-comparisons}.

Using Ashtekar's formulation, we can compare three levels of
hyperbolicity in the same interface (same fundamental variables)
as we describe next.

\subsection{Hyperbolic formulations in the Ashtekar formulations}
We present here our fundamental dynamical equations.
Our notations and a more detailed review are presented in the
Appendix \ref{appA}, but we repeat them here if necessary.

The new basic variables are
the densitized inverse triad, $\tilde{E}^i_a$, and the
SO(3,C) self-dual connection, ${\cal A}^a_i$, where the indices
$i,j,\cdots$ indicate the 3-spacetime, and
$a,b,\cdots$ are for SO(3) space.
The total four-dimensional spacetime is described together with the
gauge variables
$\null \! \mathop {\vphantom {N}\smash N}\limits ^{}_{^\sim}\!\null
, N^i, {\cal A}^a_0$, which we call the densitized lapse
function, shift vector and the triad lapse function.
The system has three constraint equations,
\begin{eqnarray}
{\cal C}^{\rm ASH}_{H} &:=&
   (i/2)\epsilon^{ab}{}_c \,
\tilde{E}^i_{a} \tilde{E}^j_{b} F_{ij}^{c}
     \approx 0, \label{const-ham} \\
{\cal C}^{\rm ASH}_{M i} &:=&
    -F^a_{ij} \tilde{E}^j_{a} \approx 0, \label{const-mom}\\
{\cal C}^{\rm ASH}_{Ga} &:=&  {\cal D}_i \tilde{E}^i_{a}
   \approx 0,  \label{const-g}
\end{eqnarray}
which are called the Hamiltonian, momentum, and Gauss constraints equation,
respectively.
The dynamical equations for a set of
$(\tilde{E}^i_a, {\cal A}^a_i)$ are
\begin{eqnarray}
\partial_t {\tilde{E}^i_a}
&=&-i{\cal D}_j( \epsilon^{cb}{}_a  \, \null \!
\mathop {\vphantom {N}\smash N}\limits ^{}_{^\sim}\!\null
\tilde{E}^j_{c}
\tilde{E}^i_{b})
+2{\cal D}_j(N^{[j}\tilde{E}^{i]}_{a})
+i{\cal A}^b_{0} \epsilon_{ab}{}^c  \, \tilde{E}^i_c,  \label{eq-E}
\\
\partial_t {\cal A}^a_{i} &=&
-i \epsilon^{ab}{}_c  \,
\null \! \mathop {\vphantom {N}\smash N}\limits ^{}_{^\sim}\!\null
\tilde{E}^j_{b} F_{ij}^{c}
+N^j F^a_{ji} +{\cal D}_i{\cal A}^a_{0},
\label{eq-A}
\end{eqnarray}
where
${F}^a_{ij}
:=
2 \partial_{[i} {\cal A}^a_{j]}
   - i \epsilon^{a}{}_{bc} \, {\cal A}^b_i{\cal A}^c_j
$
is the curvature 2-form.

We have to consider the reality conditions when we use this
formalism to describe the classical Lorentzian spacetime.
As we review in \S \ref{appendixreality}, the metric will remain on
its real-valued constraint surface during time evolution
   automatically if we prepare initial data which satisfies the
reality condition. More practically, we further require that triad is
real-valued.  But again this reality condition appears as a gauge
restriction on ${\cal A}^a_0$, (\ref{s-reality2-final}), which
can be imposed at every time step. In our actual simulation, we
prepare our initial data using the standard ADM approach, so that
we have no difficulties in maintaining  these reality conditions.

The set of
dynamical equations (\ref{eq-E}) and (\ref{eq-A}) [hereafter we call
these the {\it original} equations] does have
a weakly hyperbolic form \cite{YS-IJMPD}, so that we regard
the mathematical structure of the original equations as one step
advanced from the standard ADM.
Further, we can construct higher levels of
hyperbolic systems by restricting the gauge condition and/or
by adding constraint terms,
${\cal C}^{\rm ASH}_{H}, {\cal C}^{\rm ASH}_{Mi}$ and
${\cal C}^{\rm ASH}_{Ga}$, to the original equations
\cite{YS-IJMPD}.  We extract only the final expressions here.

In order to obtain a
symmetric hyperbolic system
\footnote{
Iriondo et al\cite{Iriondo} presented
a symmetric hyperbolic expression
in a different form.
The differences between ours and theirs are discussed in
\cite{YS-IJMPD,YShypPRL}}, we add constraint terms to
the right-hand-side of
(\ref{eq-E}) and (\ref{eq-A}).  The adjusted dynamical equations,
\begin{eqnarray}
\partial_t {\tilde{E}^i_a}
&=&-i{\cal D}_j( \epsilon^{cb}{}_a  \, \null \!
\mathop {\vphantom {N}\smash N}\limits ^{}_{^\sim}\!\null
\tilde{E}^j_{c}
\tilde{E}^i_{b})
+2{\cal D}_j(N^{[j}\tilde{E}^{i]}_{a})
+i{\cal A}^b_{0} \epsilon_{ab}{}^c  \, \tilde{E}^i_c
+P^i{}_{ab}  \, {\cal C}^{\rm ASH}_G{}^b,  \label{eqE2} \\
&~& {\rm where} \qquad P^i{}_{ab} \equiv
N^i \delta_{ab}+i\null \! \mathop {\vphantom {N}\smash N}
\limits ^{}_{^\sim}\!\null  \epsilon_{ab}{}^{c}\tilde{E}^i_c,
\nonumber
\\
\partial_t {\cal A}^a_{i} &=&
-i \epsilon^{ab}{}_c  \,
\null \! \mathop {\vphantom {N}\smash N}\limits ^{}_{^\sim}\!\null
\tilde{E}^j_{b} F_{ij}^{c}
+N^j F^a_{ji} +{\cal D}_i{\cal A}^a_{0}
+Q^a_i {\cal C}^{\rm ASH}_H
+R_i{}^{ja}  \, {\cal C}^{\rm ASH}_{Mj}, \label{eqA2}
\\
&~& {\rm where} \qquad Q^a_{i} \equiv
e^{-2}
\null \! \mathop {\vphantom {N}\smash N}\limits ^{}_{^\sim}\!\null
\tilde{E}^a_i, \qquad
R_i{}^{ja} \equiv
ie^{-2}
\null \! \mathop {\vphantom {N}\smash N}\limits ^{}_{^\sim}\!\null
\epsilon^{ac}{}_b \tilde{E}^b_i \tilde{E}^j_c
\nonumber
\end{eqnarray}
form a symmetric hyperbolicity if we further
require the gauge conditions,
\begin{equation}
{\cal A}^a_0={\cal A}^a_i N^i, \qquad \partial_i N =0.
\label{symhypgauge}
\end{equation}
We remark that the adjusted coefficients,
$P^i{}_{ab}, Q^a_i, R_i{}^{ja}$, for
constructing the symmetric  hyperbolic system are uniquely determined,
and
there are no other additional terms (say, no ${\cal C}^{\rm ASH}_H,
{\cal C}^{\rm ASH}_M$ for $\partial_t \tilde{E}^i_a$,
no ${\cal C}^{\rm ASH}_G$ for $\partial_t {\cal A}^a_i$)
\cite{YS-IJMPD}.
The gauge conditions, (\ref{symhypgauge}), are
consequences of the consistency with (triad) reality conditions.

We can also construct a strongly (or diagonalizable) hyperbolic system
by restricting to a gauge
$N^l \neq 0, \pm N \sqrt{\gamma^{ll}}$
(where $\gamma^{ll}$ is the three-metric and we do not sum indices here)
for the original equations (\ref{eq-E}), (\ref{eq-A}).
Or we can also construct from the adjusted equations,
(\ref{eqE2}) and (\ref{eqA2}),
   together with the gauge condition
\begin{equation}
{\cal A}^a_0={\cal A}^a_i N^i.  \label{diagohypgauge}
\end{equation}
As for the strongly hyperbolic system, we hereafter take the latter
expression.

In Table \ref{eqmtable}, we have summarized the equations to
be used throughout the remainder of this article.

\begin{table}[h]
\begin{center}
\begin{tabular}{cl||l|l|c}
& system & variables & Eqs of motion & remark
\\
\hline \hline
& ADM & ($\gamma_{ij}, K_{ij}$)  & (\ref{hatten1}), (\ref{hatten2}) &
``standard ADM"
\\  \hline
I & Ashtekar (weakly hyp.) & ($\tilde{E}^i_a, {\cal A}^a_i$) &
(\ref{eq-E}), (\ref{eq-A}) (original) & ``original Ashtekar"
\\ \hline
II & Ashtekar (strongly hyp.) & ($\tilde{E}^i_a, {\cal A}^a_i$) &
(\ref{eqE2}), (\ref{eqA2}) (adjusted)&
(\ref{diagohypgauge}) required
\\ \hline
III & Ashtekar (symmetric hyp.) & ($\tilde{E}^i_a, {\cal A}^a_i$) &
(\ref{eqE2}), (\ref{eqA2}) (adjusted)&
(\ref{symhypgauge}) required
\\ \hline
(in \S \ref{Results2}) & Ashtekar (adjusted) &
($\tilde{E}^i_a, {\cal A}^a_i$) &
(\ref{eqE3}), (\ref{eqA3}) (adjusted with $\kappa$)&
\\
\end{tabular}

\caption{List of systems that we compare in this article.  }
\label{eqmtable}
\end{center}
\end{table}

\section{Numerical method} \label{sec3}
\subsection{Overview}
We coded up the program so as to compare the evolutions of spacetime
with different set of dynamical equations but with the common conditions:
the same initial data, the same boundary conditions, the same slicing
condition and the same evolution scheme.

We consider the plane symmetric vacuum spacetime without cosmological
constant.  This spacetime has the true freedom of gravitational waves
of two polarized ($+$ and $\times$) modes.
We apply the periodic boundary conditions to remove any difficulties
caused by numerical treatment of the boundary conditions.
The initial data are given by solving constraint equations in ADM
variables, using the standard conformal approach by York and
O'Murchadha \cite{yorkomurchadha}.
When we use Ashtekar's variables for evolution,
we transform the ADM initial data in terms of Ashtekar's variables.
The results are analyzed by
monitoring the violation of the constraint equations which are
expressed using the same (or transformed if necessary) variables.

We describe our procedures in the following subsections in detail.
\subsection{metric and the initial data construction}\label{initdata}
We consider the plane symmetric metric,
\begin{equation}
ds^2 =
(-N^2+N_xN^x)dt^2+2 N_x dxdt + \gamma_{xx} dx^2 +
\gamma_{yy} dy^2 + \gamma_{zz} dz^2 + 2 \gamma_{yz} dydz
\label{metric}
\end{equation}
where the components are the function of
$N(x,t), N_x(x,t), \gamma_{xx}(x,t), \gamma_{yy}(x,t),
\gamma_{zz}(x,t), \gamma_{yz}(x,t)$. $N$ and $N^x$ are called the lapse
function and the shift vector.

We prepare our initial data by solving the ADM constraint equations,
(\ref{conhamilt}) and (\ref{conmoment}), using
the conformal approach \cite{yorkomurchadha}.
Since we consider only the vacuum spacetime, the input quantities are
the initial guess of the 3-metric
$\hat{\gamma}_{ij}$, the trace part of the extrinsic curvature
$~\mbox{tr} K$,
and the transverse traceless part of
the extrinsic curvature $\hat{A}_{TT}$. For simplicity, we impose
$\hat{A}_{TT}=0$ and
$
\mbox{tr} K = K_0
$ (constant).
The Hamiltonian constraint, then, becomes an equation
for the conformal factor,
$\psi$:
\begin{equation}
8\hat{\Delta}\psi :=
8 {1 \over \sqrt{\hat{\gamma}} } \partial_i
( \hat{\gamma}^{ij} \sqrt{\hat{\gamma}}
\partial_j \psi ) =
\hat{R}\psi
+ {2 \over 3} ( K_0)^2 \psi^5,
\label{bchamilt_vac}
\end{equation}
where $\hat{\gamma}=\det \hat{\gamma}_{ij}$.
The momentum constraint is automatically satisfied by assumption.
The initial dynamical quantities $\gamma_{ij}, \, K_{ij}$ are given by
the conformal transformation,
\begin{equation}
\gamma_{ij} = \psi^4 \hat{\gamma}_{ij},  \qquad
K_{ij} =
{1 \over 3}\psi^4\hat{\gamma}_{ij} K_0.
\end{equation}

We solve (\ref{bchamilt_vac})
under the periodic boundary conditions using the
incomplete Cholesky conjugate gradient (ICCG) method.

We should remark here that we have to assume non-zero
$K_0$ for a model of gravitational pulse waves under the periodic
boundary conditions in this plane symmetric spacetime.
This can be seen as follows.
Suppose we set $K_0=0$. {}From (\ref{bchamilt_vac}), we get
\begin{equation}
\partial_x \psi = {1 \over \sqrt{\gamma}g^{xx}} \int
\sqrt{\gamma} R \psi dx. \label{integeratedHam}
\end{equation}
If we set the boundary as $x=[A,B]$ and impose the periodic
boundary conditions, then eq.
(\ref{integeratedHam}) becomes
\begin{equation}
\partial_x \psi|_{x=A}-\partial_x \psi|_{x=B} =
\left( {1 \over \sqrt{\gamma}g^{xx}} \right)_{x=A=B}
\int_B^A \sqrt{\gamma} R \psi dx.
\end{equation}
However when there
exist a gravitational wave pulse which produces $R\neq 0$ in the region,
this equation gives $\partial_x \psi|_{x=A}=\partial_x \psi|_{x=B}$,
which is  inconsistent with the periodic boundary conditions.
Therefore we need to assume non-zero $K_0$ in order to compensate
the curvature which is produced by the pulse waves.

Actually the trace of the extrinsic curvature appears only in the
quadratic form, so we can interpret that our (background)
spacetime is either
expanding, $K_0 <0$, or contracting, $K_0 >0 $.
However, this fact indicates that there is no known exact solution to
compare with.
If the background  aspacetime is allowed to be flat ($K_0=0$),
then we know there is a series of exact solutions
which describes a collision of plane gravitational waves
which were
originally found by Szekerez and Khan-Penrose \cite{colwave}.
The formation of a curvature singularity after such colliding
waves is known to be generic, but that is not generalized to the
expanding background (as discussed using numerical simulations
\cite{cenmatz,ShinkaiMaeda}).

We can set two different modes of gravitational waves.
One is the $+$-mode waves, which is given by setting a
conformal guess metric as (in a matrix form)
\begin{equation}
\hat{\gamma}_{ij}=
\left(\matrix{
1&0&0\cr
{\it sym.} &1+ a \exp ( - b (x-c)^2)&0\cr
{\it sym.} & {\it sym.} & 1- a \exp ( - b (x-c)^2)}
\right)
\label{plusmetric}
\end{equation}
where $a,b,c$ are parameters. The other is the
$\times$-mode waves, given by
\begin{equation}
\hat{\gamma}_{ij}=
\left(\matrix{
1&0&0\cr
{\it sym.}  &1 &a \exp ( - b (x-c)^2) \cr
{\it sym.}  & {\it sym.} & 1}
\right)
\label{crossmetric}
\end{equation}
where $a,b,c$ are parameters again.
Both cases, we expect non-linear behavior when wave's curvature
becomes quite large compared to the background.
In the collision of a $+$-mode wave and a $\times$-mode wave, we also
expect to see the mode-mixing phenomena which is known as gravitational
Faraday effect.
These effects are confirmed in our numerical simulations.

\subsection{Transformation of variables: From ADM to Ashtekar} \label{ADM2ASH}
We need to transform the dynamical variables on the initial data
when we evolve them in the connection variables.  We list
the procedure to
obtain $(\tilde{E}^i_a, {\cal A}^a_i)$ from $(\gamma_{ij}, K_{ij})$.
This procedure is used also when we evaluate the constraints,
   ${\cal C}^{\rm ASH}_{H}, {\cal C}^{\rm ASH}_{Mi},
{\cal C}^{\rm ASH}_{Ga}$ for the data evolved using ADM variables.

{}From the three-metric $\gamma_{ij}$ to $\tilde{E}^i_a$:
\begin{enumerate}
\item Define the triad $E^a_i$ corresponding to the
three-metric $\gamma_{ij}$. We take
\begin{equation}
E^a_i=
\left [\begin {array}{ccc}
E^1_x & E^1_y & E^1_z \\
E^2_x & E^2_y & E^2_z \\
E^3_x & E^3_y & E^3_z
\end {array}\right ] = \left [\begin {array}{ccc}
\sqrt{\gamma_{xx}} & 0 & 0 \\
0 & e_{22} & e_{23} \\
0 & e_{32} & e_{33} \\
\end {array}\right ]. \label{plane-triad}
   \end{equation}
and set simply $e_{23}=e_{32}$.
The relation between the metric and the triad becomes
\begin{equation}
e_{22}^2+e_{33}^2=\gamma_{yy}, \qquad e_{23}^2+e_{33}^2=\gamma_{zz},\qquad
(e_{22}+e_{33})e_{23}=\gamma_{yz}.
\end{equation}
For the case of $+$-mode waves, we define naturally,
$e_{22}=\sqrt{\gamma_{yy}}, e_{33}=\sqrt{\gamma_{zz}}, e_{23}=0$.
For $\times$-mode waves, we also take a natural set of definitions,
$e_{22}=e_{33}=[(\gamma_{yy}+(\gamma_{yy}^2-\gamma_{yz}^2)^{1/2})/2]^{1/2}$ and
$e_{23}=\gamma_{yz}/2e_{22}$ which are given by solving
$e_{22}^2+e_{33}^2=\gamma_{yy}$ and $2e_{22}e_{23}=\gamma_{yz}$.
\item obtain the inverse triad $E^i_a$ from triad $E^a_i$.
\item calculate the density, $e$,  as $e=\det E^a_i$.
\item obtain the densitized triad, $\tilde{E}^i_a = e E^i_a$.
\end{enumerate}

{}From three-metric $(\gamma_{ij}, K_{ij})$ to ${\cal A}^a_i$:
\begin{enumerate}
\item prepare the
triad $E^a_i$ and its
inverse $E^i_a$.
\item calculate the connection 1-form
$\omega^{bc}_i=E^{b\mu} \nabla_i E^c_\mu$.  This is expressed
only using partial derivatives as
\footnote{This is from the definitions,
$\omega^{bc}_i :=
E^{j b} \nabla_i E^c_j
$ and $
\omega^{abc} :=  E^{j a}\omega^{bc}_j
$, and a relation
$$
3\omega^{[abc]}
-2\omega^{[bc]a} =
\omega^{a[bc]}
+\omega^{b[ca]}
+\omega^{c[ab]}
-\omega^{abc}
+\omega^{cba}
=
\omega^{abc}.
$$
Using the densitized triad, eq. (\ref{eq310}) can be also expressed as
$$
\omega^{bc}_i =
   {2\over e^2}\tilde{E}^{jb} (\partial_{[i} \tilde{E}^c_{j]})
+{1\over e^4}\tilde{E}^{jb} \tilde{E}^c_i\tilde{E}^a_k(\partial_j
\tilde{E}^k_a)
+{1\over 4e^4}\tilde{E}_{i a}\tilde{E}^{kb} \tilde{E}^j_c (\partial_j
\tilde{E}^a_k),
\quad \mbox{taking }[bc].
$$
}
\begin{equation}
\omega^{bc}_i =
E^{j b} \partial_{[i} E^c_{j]}
-E_{i d}E^{k b}E^{j c} \partial_{[k} E^d_{j]}
+E^{j c} \partial_{[j} E^b_{i]}. \label{eq310}
\end{equation}
\item
$
{\cal A}^a_i=-K_{ij}E^{ja}-{i \over 2}
\epsilon^a{}_{bc}\omega^{bc}_i.
$
\end{enumerate}

\subsection{Transformation of variables: From Ashtekar to ADM} \label{ASH2ADM}
In contrast to the previous transformation, we also need to obtain
($\gamma_{ij}, K_{ij}$) from ($\tilde{E}^i_a, {\cal A}^a_i$)
when we evaluate
the metric output or ADM constraints when we evolve the spacetime using
connection variables.  This process is only required at an evaluation times,
not required at every time step (unless we use the gauge condition which is
primarily defined using ADM quantities).

{}From densitized inverse triad $\tilde{E}^i_a$ to
three-metric $\gamma_{ij}$:
\begin{enumerate}
\item calculate the density $e$ as $e = ({\det \tilde{E}^i_a})^{1/2}$.
\item get the three inverse metric as
$\gamma^{ij}=\tilde{E}^i_a\tilde{E}^j_a / e^2$.
\item obtain $\gamma_{ij}$.
\end{enumerate}

{}From $(\tilde{E}^i_a, {\cal A}^a_i)$ to the extrinsic curvature $K_{ij}$:
\begin{enumerate}
\item prepare the un-densitized inverse triad, $E^i_a = \tilde{E}^i_a / e$.
\item prepare triad $E_i^a$.
\item calculate the connection 1-form $\epsilon^a{}_{bc}\omega^{bc}_i$.
\item calculate $Z^a_i$, which is defined as
\footnote{
This is from the original definition of ${\cal A}^a_i$,
${\cal A}^a_{i}
:= \omega^{0a}_i - ({i / 2}) \epsilon^a{}_{bc} \, \omega^{bc}_i.
$}
$Z^a_i := -{\cal A}^a_i+{i \over 2} \epsilon^a{}_{bc}
\omega^{bc}_i (=K_{ij}E^{ja})
$, and get
$K_{ij}=Z^a_iE_{ja}$.
\end{enumerate}

\subsection{Gauge conditions}
We evolve the initial data with different evolution equations and compare
its accuracy/stability.
As we summarized in Table \ref{eqmtable},
we will compare time evolutions between ADM
and Ashtekar (of the original system I) in \S \ref{ADMvsASH}, and
three of Ashtekar's systems (I, II and III: weakly, stronly and symmetric)
in \S \ref{ASH123}.
We, then, consider alternative system (adjusted $\kappa$ system)
in \S \ref{Results2}.

Here
we comment again on our choice of the slicing (gauge) condition.
As for the primary tests of this subject, we apply the simplest
slicing conditions we can take.  That is,
\begin{enumerate}
\item[(1)] the simplest geodesic slicing condition
for the lapse function,
\item[(2)] the simplest zero shift vector $N^x=0$, and
\item[(3)] the natural choice of triad lapse function
${\cal A}^a_0={\cal A}^a_iN^i \, [=0 \mbox{~if~} N^x=0$,
which is suggested from
(\ref{symhypgauge}) or (\ref{diagohypgauge})].
\end{enumerate}
However, in the Ashtekar formalism, the densitized
lapse function
$\null \! \mathop {\vphantom {N}\smash N}\limits ^{}_{^\sim}\!\null $
is the fundamental gauge quantity (rather than $N$).
Therefore we try two
conditions for the lapse,
\begin{enumerate}
\item[(1a)] the standard geodesic slicing
condition $N=1$, which
will be transformed to
$\null \! \mathop {\vphantom {N}\smash N}\limits ^{}_{^\sim}\!\null
= 1/ e$ when we apply this condition
in Ashtekar's evolution system,
and
\item[(1b)] the densitized geodesic slicing condition
$\null \! \mathop {\vphantom {N}\smash N}\limits ^{}_{^\sim}\!\null =1$,
which will be
transformed to $N=e$ when we evolve the system using ADM equations.
\end{enumerate}
In practice, such a transformation using the density $e$ will not
guarantee that the Courant condition
holds if we fix the time evolution step $\Delta t$
\footnote{We here remind the reader of the stability condition,
$N \Delta t \leq \Delta x$
for a standard forward-time centered-space (FTCS)
scheme for a simple wave equation, in a
   ($\Delta t$, $\Delta x$)-spaced numerical grid.  Note that
this condition will be changed due to the choice of
the evolution scheme and the equations of the system.
}.  Therefore we need to rescale the
transformed lapse [
$\null \! \mathop {\vphantom {N}\smash N}\limits ^{}_{^\sim}\!\null $
in (1a), $N$ in (1b)] so that
it has a maximum
value of unity, in order to keep our evolution system stable.

If we apply the standard geodesic slice, then we can compare
the weakly hyperbolic
system with the symmetric hyperbolic one. Similarly if we
apply the densitized geodesic
slice, then we can compare the (original) weakly hyperbolic
system with the strongly hyperbolic one.

\subsection{Time integrating scheme}
We applied two second order evolution schemes, and confirmed that they
give us nearly identical results.

One is the so-called iterative
Crank-Nicholson scheme (cf. \cite{TeukolskyCrankNicholson}),
which is now becoming the standard in the numerical relativity community.
Suppose we have a dynamical equation in the form
\begin{equation}
\partial_t u(x,t) = f(u(x,t), \partial_x u(x,t)).
\end{equation}
Then the scheme for updating $u$ at a point $x$ from $t$ to $t+\Delta t$
consists of the following steps.
(1) use data on $t=t$
for right-hand-side and update $u(x,t)$ for $\Delta t$ step as
$\tilde{u}(x,t+ \Delta t)$,
\begin{equation}
{\tilde{u}(x,t+ \Delta t) - u(x, t) \over \Delta t}=
f( u(x, t), \partial_x u(x, t)).
\label{firststep}
\end{equation}
(2) take the average of $u(x,t)$ and $\tilde{u}(x,t+ \Delta t)$,
and let it represent a half-step value, (say $v(x,t+\Delta t/2 )$).
(3) update $u(x,t)$ for $\Delta t$ step again using $\hat{u}(x,t+\Delta t/2 )$
in the argument of the right-hand-side,
\begin{equation}
{\tilde{u}(x,t+ \Delta t) - u(x, t) \over \Delta t}=
f( \hat{u}(x, t+\Delta t/2 ), \partial_x \hat{u}(x, t+\Delta t/2 )).
\end{equation}
(4) perform the above (2) and (3) steps once more (we suppose two-iteration
Crank-Nicholson scheme), and take
$\tilde{u}(x,t+ \Delta t)$ to be the evolved quantity.

The other scheme we applied is the
Brailovskaya integration scheme, which is
a second order predictor-corrector method \cite{bernstein}
and rather easy to code.  The first step (predictor step)
is the same as (\ref{firststep}),
and the second step (corrector step)
simply switches the right-hand-side
using the updated
$\tilde{u}(x,t+ \Delta t)$ to be
\begin{equation}
{u(x,t+ \Delta t) - u(x, t) \over \Delta t}=
f(\tilde{u}(x,t+ \Delta t), \partial_x \tilde{u}(x,t+ \Delta t)).
\end{equation}
Note that all derivatives here in the right-hand-side are
assumed to use
central difference.

The latter scheme is quite simple, but gives us
reasonably accurate and
stable evolutions for our problems.
We confirmed
that both give us nearly identical evolutions (which will be
shown in Fig.\ref{admash} (b)),  but the Brailovskaya
method requires less
computational time.

\subsection{Checking the constraints}
We compare the violation of the constraint equations during
the time evolution.
We have
ADM constraint equations, ${\cal C}^{\rm ADM}_{H}$ and
${\cal C}^{\rm ADM}_{Mi}$
[(\ref{conhamilt}) and (\ref{conmoment})], and also
Ashtekar's constraint equations, ${\cal C}^{\rm ASH}_{H},
{\cal C}^{\rm ASH}_{Mi}$
and ${\cal C}^{\rm ASH}_{Ga}$
[(\ref{const-ham}),
(\ref{const-mom}) and
(\ref{const-g}), respectively].
By means of the transformation between $(\gamma_{ij}, K_{ij})$
and $(\tilde{E}^i_a, {\cal A}^a_i)$,
we can evaluate ADM constraints even if we evolved the system
using Ashtekar's variables and vice versa.

We measure the violation of a constraint by its
(i) maximum, $\mbox{\rm max}_x |{\cal C}(x)|$,
(ii) L1 norm,  $ (\sum_{x=1}^{n_x}  {\cal C}(x))/ n_x$, and
(iii) L2 norm,  $ (\sum_{x=1}^{n_x} |{\cal C}(x)|^2/ n_x)^{1/2}$,
where $n_x$ is the number of grid points.
When we compare them during the evolution, we measure them at the
same proper time, $\tau$, for the two different evolution systems.
The proper time is defined locally as $d\tau = N dt$, which is also
$d\tau = e
\null \! \mathop {\vphantom {N}\smash N}\limits ^{}_{^\sim}\!\null dt$,
but here we apply its averaged value on the whole $t=$constant
surface, (say $\langle N \rangle = (1/n_x)\sum_{x=1}^{n_x} N$),
\begin{equation}
\tau=\int^t_0
\langle N \rangle \, dt, \qquad \mbox{ or } \qquad
\tau=\int^t_0 \langle e
\null \! \mathop {\vphantom {N}\smash N}\limits ^{}_{^\sim}\!\null
   \rangle \, dt,
\end{equation}
to characterize the ``time" of evolution.

The numerical code passed convergence tests, and the results shown in
this article are all obtained with acceptable accuracy.
In Fig.\ref{conv_test}, we show a result of convergence tests.
We show the convergence behaviour of our initial data solver
in Fig.\ref{conv_test}(a), together with the convergence
behaviour of the evolution codes both for
ADM and Ashtekar variables in Fig.\ref{conv_test}(b) and (c).
We plotted the residual of the Hamiltonian constraint solver
when it was minimized, and the L2 norm of ${\cal C}^{\rm ADM}_H$
and  ${\cal C}^{\rm ASH}_H$ for
the evolution of $+$-mode
single pulse wave.
(The model is described in the next section.)
We can see all errors are diminished by finer resolutions.
The order of the convergence\footnote{
Here we used the definition of the order of convergence following
Bona {\it et. al.} \cite{9804052}.} is 1.98 to 2.01 for the initial
data solver, and at best 1.96 (e.g. at $\tau=0.5$) for the evolution code.

All the results we present in this article are
obtained using 401 grid points for the range $x=[-5,+5]$;
that is gravitational waves traverse the entire numerical region
in proper time 10 if the background expansion $K_0$ is close to zero.
We use the Courant number $\nu = \Delta t / \Delta x =0.2$.

We coded all our  fundamental quantities (metric, gauge
variables, $\cdots$) as complex, but we observed that
the evolution from our initial data never violate its
metric reality conditions. Due to the gauge condition for ${\cal A}^a_0$,
(\ref{s-reality2-final}), we also confirmed that
our evolutions preserve
the triad reality condition.

\section{Experiments 1: Differences between hyperbolicities}
\label{Results1}
In this section, we examine the accuracy/stability of the
numerical evolutions comparing the different hyperbolic systems.
We begin showing how the evolution of Ashtekar's equations
look, comparing with those of the  ADM equations.
\subsection{ADM vs Ashtekar}
\label{ADMvsASH}
We start by describing our model, plane
wave propagation in expanding/collapsing spacetime.
We prepare the initial data with one or two gravitational
pulse waves in our numerical region.  The pulses then start propagating
in both $\pm x$ directions with the light speed, and appear on the
other side of the numerical region
due to the periodic boundary condition.  When the pulses collide, then
the amplitude seems simply to double, as they are superposed, and
the pulses keep traveling in their original propagation direction.
That is, we observe something like solitonic wave pulse
propagation.

As we mentioned in \S \ref{initdata}, we have to assume our background
not to be flat, therefore there is no exact solutions.
The reader might think that
if we set $|\mbox{tr~}K |$ to be small and pulse wave shapes
to be quite sharp
then our simulations will be close to the analytic
colliding plane wave solutions which produce the
curvature singularity.
However, from the numerical side, these two
requirements are contradictory (e.g. sharp wave input
produces large curvature
which should be compensated by $|\mbox{tr~}K |$ in order to
construct our initial data). Thus it is not so surprising that
our waves propagate like solitons, not forming a singularity.

In Fig.\ref{admash} (a), we plot
an image of wave propagation (metric component $g_{yy}$) up to $\tau=10$,
of $+$-mode pulse waves initially located at $x=\pm 2.5$.
We took a small negative $K_0$, so that the background spacetime is
slowly expanding.

Fig.\ref{admash} (b), then, tells us that
our ADM evolution code and Ashtekar's variable code give us
identical evolutions.  We plotted a snapshot of $g_{yy}$
on the initial data (which is common to all models here), and
its snapshot at $\tau=10.0$.  The fact that all four lines
(ADM/Ashtekar, of their Brailovskaya/Crank-Nicholson evolution
schemes) overlapped clearly indicates that we are showing exact
evolutions.

We also plotted a typical evolution of
the fundamental dynamical
quantities $\tilde{E}^y_2$ and ${\cal A}^2_y$
in Figs.\ref{admash} (c) and (d).

We next compare constraint violations by Ashtekar's equation with that of ADM.
In Fig.\ref{admash2}, we plot the L2 norm of ${\cal C}_H^{\rm ASH}$
and ${\cal C}_H^{\rm ADM}$.
We see that ADM
evolution shows less violation in measuring ${\cal C}_H^{\rm ADM}$,
and the Ashtekar
evolution shows less violation in measuring ${\cal C}_H^{\rm ASH}$.
The magnitudes of these violations are similar. Thus, we believe,
these violations are within the numerical truncation errors in the process of
numerical transformation of variables (ADM to Ashtekar/ Ashtekar to ADM), and
therefore  it is not appropriate to conclude here which
formulation is better.

As the reader may guess, the violations of constraints reduce
if the background spacetime is expanding ($K_0<0$). Therefore we will use the
collapsing background spacetime ($K_0>0$) hereafter for presentations,
with the expectation of having more non-linear effects; however this
direction also stops the evolution after the finite time.
(e.g. for the flat initial data with $K_0=+0.025$, the spacetime will
collapse to zero volume around $t=60$.)

\subsection{Comparison between hyperbolicities}\label{ASH123}

We here present our comparisons of the accuracy and/or stability between the
different hyperbolicities.
Since all examples we show in this section are not the case of unstable
evolution (no exponential growth of constraint violation),  our experiments
can be said as the comparisons of the accuracy of the evolution, 
conservatively.

We first compare the (original) weakly hyperbolic system [system I in
Table \ref{eqmtable}] with the strongly hyperbolic system  [system II in
Table \ref{eqmtable}]. This comparison can be done under the
densitized geodesic slicing condition,
$\null \! \mathop {\vphantom {N}\smash N}\limits ^{}_{^\sim}\!\null =1$.
We prepare two initial gravitational pulses (both
$+$-$+$ or $\times$-$\times$ modes)
and take our background spacetime to be collapsing ($K_0 > 0$).
In Fig.\ref{hikaku_Ne}, we show the constraint errors,
${\cal C}^{\rm ASH}_H$ and ${\cal C}^{\rm ASH}_M$.
In both two situations, we observe
that the strongly hyperbolic system has slightly improved the violation
of the constraints, but we can not see the orders of magnitude differences.

Similarly,
we next compare the (original) weakly hyperbolic system
[system I in
Table \ref{eqmtable}] with the symmetric hyperbolic system
[system III in
Table \ref{eqmtable}]. This comparison can be done under
the standard geodesic slicing condition, $N=1$.  We repeat the same
experiments as above and show plots in Fig.\ref{hikaku_N1}.
We again see that the symmetric hyperbolic system slightly
improves the situation, but not so drastically.

{}From both Figs. \ref{hikaku_Ne} and \ref{hikaku_N1}, we see that the
strongly and symmetric hyperbolic systems produce less violation
of constraints than the original weakly hyperbolic system.
Therefore one conclusion is that
adjusting the equation of motion with constraint terms
{\it does} definitely make the system accurate.
However the constraint violation remains the same order of magnitude.

 From each figures, we may conclude that
higher level hyperbolic system
gives us slightly accurate evolutions.
However, if we evaluate the magnitude of L2 norms, then
we also conclude that there is no measurable differences between
strongly and symmetric hyperbolicities.
This last fact will be supported more affirmatively in the next
experiment.

\section{Experiments 2: Another way to control the accuracy/stability}
\label{Results2}

The results we have presented in the previous section indicate that
both strongly and symmetric hyperbolic systems show better performance
than the original weakly hyperbolic system.
These systems are obtained by adding constraint terms
(or ``adjusted" terms) to the right-hand-side of the original equations,
(\ref{eq-E}) and (\ref{eq-A}).
In this section, we report
on simple experiments in changing the magnitude of
the multipliers of such adjusted terms.

We consider the following system, where the
equations of motion are adjusted in the same way as before,
but with a real-valued constant multiplier $\kappa$:
\begin{eqnarray}
\partial_t {\tilde{E}^i_a}
&=&-i{\cal D}_j( \epsilon^{cb}{}_a  \, \null \!
\mathop {\vphantom {N}\smash N}\limits ^{}_{^\sim}\!\null
\tilde{E}^j_{c}
\tilde{E}^i_{b})
+2{\cal D}_j(N^{[j}\tilde{E}^{i]}_{a})
+i{\cal A}^b_{0} \epsilon_{ab}{}^c  \, \tilde{E}^i_c
+ \kappa P^i{}_{ab}  \, {\cal C}^{\rm ASH}_G{}^b,  \label{eqE3} \\
&~& {\rm where} \qquad P^i{}_{ab} \equiv
N^i \delta_{ab}+i\null \! \mathop {\vphantom {N}\smash N}
\limits ^{}_{^\sim}\!\null  \epsilon_{ab}{}^{c}\tilde{E}^i_c,
\nonumber
\\
\partial_t {\cal A}^a_{i} &=&
-i \epsilon^{ab}{}_c  \,
\null \! \mathop {\vphantom {N}\smash N}\limits ^{}_{^\sim}\!\null
\tilde{E}^j_{b} F_{ij}^{c}
+N^j F^a_{ji} +{\cal D}_i{\cal A}^a_{0}
+\kappa  Q^a_i {\cal C}^{\rm ASH}_H
+\kappa  R_i{}^{ja}  \, {\cal C}^{\rm ASH}_{Mj}, \label{eqA3}
\\
&~& {\rm where} \qquad Q^a_{i} \equiv
e^{-2}
\null \! \mathop {\vphantom {N}\smash N}\limits ^{}_{^\sim}\!\null
\tilde{E}^a_i, \qquad
R_i{}^{ja} \equiv
ie^{-2}
\null \! \mathop {\vphantom {N}\smash N}\limits ^{}_{^\sim}\!\null
\epsilon^{ac}{}_b \tilde{E}^b_i \tilde{E}^j_c.
\nonumber
\end{eqnarray}
The set of (\ref{eqE3}) and (\ref{eqA3})
becomes the original weakly hyperbolic system if $\kappa=0$,
becomes the symmetric hyperbolic system if $\kappa=1$ and
$N=const.$, and remains strongly hyperbolic systems
for other choices of $\kappa$ except $\kappa=1/2$ which only forms
  weakly hyperbolic system.
We remark again that the coefficients for constructing the symmetric
hyperbolic system are uniquely determined.

We tried the same evolutions as in the previous section for different
value of  $\kappa$.
In Fig.\ref{kappa}, we plot the L2 norm of the Hamiltonian and
momentum constraint equations, ${\cal C}_H^{\rm ASH}$
and ${\cal C}_M^{\rm ASH}$.
We checked first that $\kappa=0$ and 1 produce the same results
as those of weakly and symmetric hyperbolic systems.
What is interesting is the case of
$\kappa=2$ and 3.
These $\kappa$s
produce better performance than the symmetric
hyperbolic system, although these cases are of strongly hyperbolic
levels.
Therefore, as far as monitoring the violation of the constraints
is concerned, we may say the symmetric hyperbolic form is
{\it not always the best}.
We remark that the negative $\kappa$ will produce unstable evolution
as we plotted, while too large positive $\kappa$ will also results in
unstable evolution in the end (see $\kappa=3$ lines).

We also tried similar experiments with the vacuum Maxwell equation.
The original Maxwell equation has symmetric hyperbolicity, and
additional constraint terms
(with multiplier $\kappa$)
reduce hyperbolicity to the strong or weak level.
We show the details and a figure in
the Appendix \ref{appB}, but in short
there may be no measurable differences between strongly and symmetric
hyperbolicities.

These experiments in changing $\kappa$ are now reported in our
Paper II \cite{Paper2} more extensively.
There, we propose
a plausible explanation why such adjusted terms work for stabilizing
the system.  We introduce the idea in the Appendix \ref{appC}.
Briefly, we will conjecture
a criterion using the eigenvalues of `adjusted version'
of the constraint propagation
equations.  This analysis may explain the appearance of
phase differences between two systems, which is observed in
Figs.\ref{hikaku_Ne}, \ref{hikaku_N1} and \ref{kappa}.


\section{Discussion} \label{Summary}
Motivated by many recent proposals for hyperbolic formulations
of the Einstein equation, we studied numerically these accuracy/stability
properties with the purpose of comparing three mathematical levels
of hyperbolicity:
weakly hyperbolic, strongly hyperbolic,
and symmetric hyperbolic systems.
We apply Ashtekar's connection formulation, because this is the only
known system in which we can compare three hyperbolic levels with the same
interface.

Our numerical code demonstrates gravitational wave propagation in
plane symmetric spacetime, and we compare the ``accuracy" and/or
``stability" by monitoring the violation of the constraints.
Actually,
our experiments in \S \ref{Results1} were the comparisons of accuracy
in evolutions, while in \S \ref{Results2} we observed cases of unstable
evolutions.
By comparing with the results obtained from the weakly hyperbolic system, we
observe the strongly and symmetric hyperbolic system show
better properties with little differences between them.
Therefore we may conclude that higher levels of hyperbolic formulations
help the numerics more, though its differences are small.

However, we also found that the symmetric hyperbolic system is not always
the best for controlling accuracy or stability, by introducing a multiplier
for adjusted terms in the equations of motion.
This result suggests that a certain kind of hyperbolicity
is enough to control the violation of constraint equation.
In our case it is
the strongly hyperbolic level.
This statement is supported by an experiment in Maxwell system as we describe
in Appendix \ref{appB}.

The remaining question is:
why we can get the better performance by adding
constraint terms in the dynamical equations?
The added terms are basically {\it error} terms during the evolution
for its original dynamical equations.
Nevertheless, these terms improve the accuracy of the evolution.
We now have a plausible way to explain the reason which is discussed
in our Paper II \cite{Paper2} (its brief introduction is in Appendix \ref{appC}
in this article).
There we evaluate the eigenvalues of
the adjusted version of
the constraint propagation equations, and propose a criteria for obtaining
stability of the system.  In some cases, for example, the
decay/growth of the constraints can be predicted the signature of the
eigenvalues of the adjusted version of
the constraint propagation equations.
In \cite{Paper2}, we will
discuss this point in detail together with a numerical
demonstration of $\lambda$-systems \cite{BFHR,SY-asympAsh}.
There we also show that some choices of adjusted terms may produce unstable
evolution.


To conclude, we are glad to announce that Ashtekar's connection variables
have finally been applied in numerical simulations.
This new approach, we hope, will contribute to understanding further
of gravitational physics, and will open a new window for peeling off
interesting non-linear natures together with a step to numerical treatment
of quantum gravity.

\section*{Acknowledgements}
HS appreciates helpful comments by
Abhay Ashtekar, Jorge Pullin, Douglas Arnold, L. Samuel Finn,
and Mijan Huq, and the hospitality of the CGPG group.
He thanks Simonetta Frittelli for pointing out the reference
\cite{hern}.
We thank Bernard Kelly for careful reading the manuscript.
Numerical computations were performed using machines at CGPG.
This work was supported in part by the NSF grants PHYS95-14240,
and the Everly research funds of Penn State.
HS was supported by the Japan Society for the Promotion of Science
as a research fellow abroad.

\appendix
\section{Ashtekar's formulation of general relativity}\label{appA}
We give a brief review of the Ashtekar formulation
and the way of handling reality conditions.
This appendix is for describing our notations.

\subsection{Variables and Equations}\label{appendixAshtekar}
The key feature of  Ashtekar's formulation of general relativity
\cite{Ashtekar} is the introduction of a self-dual
connection as one of the basic dynamical variables.
Let us write
the metric $g_{\mu\nu}$ using the tetrad
$E^I_\mu$ as $g_{\mu\nu}=E^I_\mu E^J_\nu \eta_{IJ}$
\footnote{We use
$\mu,\nu=0,\cdots,3$ and
$i,j=1,\cdots,3$ as spacetime indices, while
$I,J=(0),\cdots,(3)$ and
$a,b=(1),\cdots,(3)$ are $SO(1,3)$, $SO(3)$ indices respectively.
We raise and lower
$\mu,\nu,\cdots$ by $g^{\mu\nu}$ and $g_{\mu\nu}$
(the Lorentzian metric);
$I,J,\cdots$ by $\eta^{IJ}={\rm diag}(-1,1,1,1)$ and $\eta_{IJ}$;
$i,j,\cdots$ by $\gamma^{ij}$ and $\gamma_{ij}$ (the three-metric);
$a,b,\cdots$ by $\delta^{ab}$ and $\delta_{ab}$.
We also use volume forms $\epsilon_{abc}$:
$\epsilon_{abc} \epsilon^{abc}=3!$.}.
Define its inverse, $E^\mu_I$, by
$E^\mu_I:=E^J_\nu g^{\mu\nu}\eta_{IJ}$ and we impose
$E^0_a=0$ as the gauge condition.
We define SO(3,C) self-dual and anti self-dual
connections
$
{}^{\pm\!}{\cal A}^a_{\mu}
:= \omega^{0a}_\mu \mp ({i / 2}) \epsilon^a{}_{bc} \, \omega^{bc}_\mu,
$
where $\omega^{IJ}_{\mu}$ is a spin connection 1-form (Ricci
connection), $\omega^{IJ}_{\mu}:=E^{I\nu} \nabla_\mu E^J_\nu.$
Ashtekar's plan is to use  only the self-dual part of
the connection
$^{+\!}{\cal A}^a_\mu$
and to use its spatial part $^{+\!}{\cal A}^a_i$
as a dynamical variable.
Hereafter,
we simply denote $^{+\!}{\cal A}^a_\mu$ as ${\cal A}^a_\mu$.

The lapse function, $N$, and shift vector, $N^i$, both of which we
treat as real-valued functions,
are expressed as $E^\mu_0=(1/N, -N^i/N$).
This allows us to think of
$E^\mu_0$ as a normal vector field to $\Sigma$
spanned by the condition $t=x^0=$const.,
which plays the same role as that of
Arnowitt-Deser-Misner (ADM) formulation.
Ashtekar  treated the set  ($\tilde{E}^i_{a}$, ${\cal A}^a_{i}$)
as basic dynamical variables, where
$\tilde{E}^i_{a}$ is an inverse of the densitized triad
defined by
$
\tilde{E}^i_{a}:=e E^i_{a},
$
where $e:=\det E^a_i$ is a density.
This pair forms the canonical set.

In the case of pure gravitational spacetime,
the Hilbert action takes the form
\begin{eqnarray}
S&=&\int {\rm d}^4 x
[ (\partial_t{\cal A}^a_{i}) \tilde{E}^i_{a}
+(i/2) \null \! \mathop {\vphantom {N}\smash N}\limits ^{}_{^\sim}\!\null
\tilde{E}^i_a \tilde{E}^j_b F_{ij}^{c} \epsilon^{ab}{}_{c}
-
e^2 \Lambda
\null \! \mathop {\vphantom {N}\smash N}\limits ^{}_{^\sim}\!\null
-N^i F^a_{ij} \tilde{E}^j_a
+{\cal A}^a_{0} \, {\cal D}_i \tilde{E}^i_{a} ],
   \label{action}
\end{eqnarray}
where
$\null \! \mathop {\vphantom {N}\smash N}\limits ^{}_{^\sim}\!\null
:= e^{-1}N$,
${F}^a_{\mu\nu}
:=
2 \partial_{[\mu} {\cal A}^a_{\nu]}
   - i \epsilon^{a}{}_{bc} \, {\cal A}^b_\mu{\cal A}^c_\nu
$
is the curvature 2-form,
$\Lambda$
is the cosmological constant,
${\cal D}_i \tilde{E}^j_{a}
      :=\partial_i \tilde{E}^j_{a}
-i \epsilon_{ab}{}^c  \, {\cal A}^b_{i}\tilde{E}^j_{c}$,
   and
$e^2=\det\tilde{E}^i_a
=(\det E^a_i)^2$ is defined to be
$\det\tilde{E}^i_a=
(1/6)\epsilon^{abc}
\null\!\mathop{\vphantom {\epsilon}\smash \epsilon}
\limits ^{}_{^\sim}\!\null_{ijk}\tilde{E}^i_a \tilde{E}^j_b
\tilde{E}^k_c$, where
$\epsilon_{ijk}:=\epsilon_{abc}E^a_i E^b_j E^c_k$
   and $\null\!\mathop{\vphantom {\epsilon}\smash \epsilon}
\limits ^{}_{^\sim}\!\null_{ijk}:=e^{-1}\epsilon_{ijk}$
\footnote{When $(i,j,k)=(1,2,3)$,
we have
$\epsilon_{ijk}=e$,
$\null\!\mathop{\vphantom {\epsilon}\smash \epsilon}
\limits ^{}_{^\sim}\!\null_{ijk}=1$,
$\epsilon^{ijk}=e^{-1}$, and
$\tilde{\epsilon}^{ijk}=1$.}.

Varying the action with respect to the non-dynamical variables
$\null \!
\mathop {\vphantom {N}\smash N}\limits ^{}_{^\sim}\!\null$,
$N^i$
and ${\cal A}^a_{0}$ yields the constraint equations,
\begin{eqnarray}
{\cal C}^{\rm ASH}_{H} &:=&
   (i/2)\epsilon^{ab}{}_c \,
\tilde{E}^i_{a} \tilde{E}^j_{b} F_{ij}^{c}
    -\Lambda \, \det\tilde{E}
     \approx 0, \label{c-ham} \\
{\cal C}^{\rm  ASH}_{M i} &:=&
    -F^a_{ij} \tilde{E}^j_{a} \approx 0, \label{c-mom}\\
{\cal C}^{\rm  ASH}_{Ga} &:=&  {\cal D}_i \tilde{E}^i_{a}
   \approx 0.  \label{c-g}
\end{eqnarray}
The equations of motion for the dynamical variables
($\tilde{E}^i_a$ and ${\cal A}^a_i$) are
\begin{eqnarray}
\partial_t {\tilde{E}^i_a}
&=&-i{\cal D}_j( \epsilon^{cb}{}_a  \, \null \!
\mathop {\vphantom {N}\smash N}\limits ^{}_{^\sim}\!\null
\tilde{E}^j_{c}
\tilde{E}^i_{b})
+2{\cal D}_j(N^{[j}\tilde{E}^{i]}_{a})
+i{\cal A}^b_{0} \epsilon_{ab}{}^c  \, \tilde{E}^i_c,  \label{eqE}
\\
\partial_t {\cal A}^a_{i} &=&
-i \epsilon^{ab}{}_c  \,
\null \! \mathop {\vphantom {N}\smash N}\limits ^{}_{^\sim}\!\null
\tilde{E}^j_{b} F_{ij}^{c}
+N^j F^a_{ji} +{\cal D}_i{\cal A}^a_{0}+\Lambda
\null \!\mathop {\vphantom {N}\smash N}\limits ^{}_{^\sim}\!\null
\tilde{E}^a_i,
\label{eqA}
\end{eqnarray}
\noindent
where
${\cal D}_jX^{ji}_a:=\partial_jX^{ji}_a-i
   \epsilon_{ab}{}^c {\cal A}^b_{j}X^{ji}_c,$
   for $X^{ij}_a+X^{ji}_a=0$.

\subsection{Reality conditions}\label{appendixreality}
In order to construct the metric  from the variables
$(\tilde{E}^i_a, {\cal A}^a_i,  \null \!
\mathop {\vphantom {N}\smash N}\limits ^{}_{^\sim}\!\null, N^i)$,
we first prepare the
tetrad $E^\mu_I$ as
$E^\mu_{0}=({1 / e
\null \! \mathop {\vphantom {N}\smash N}\limits ^{}_{^\sim}\!\null
}, -{N^i / e
\null \!\mathop {\vphantom {N}\smash N}\limits ^{}_{^\sim}\!\null
})$ and
$E^\mu_{a}=(0, \tilde{E}^i_{a} /e).$
Using them, we obtain the  metric $g^{\mu\nu}$ such that
$
g^{\mu\nu}:=E^\mu_{I} E^\nu_{J} \eta^{IJ}. 
$

This metric, 
in general, is not real-valued
in the Ashtekar
formulation.
To ensure that the metric is real-valued,
we need to impose real lapse and shift vectors together with
two {\it metric reality} conditions;
\begin{eqnarray}
{\rm Im} (\tilde{E}^i_a \tilde{E}^{ja} ) &=& 0, \label{w-reality1} \\
W^{ij}:= {\rm Re} (\epsilon^{abc}
\tilde{E}^k_a \tilde{E}^{(i}_b {\cal D}_k \tilde{E}^{j)}_c)
&=& 0,
\label{w-reality2-final}
\end{eqnarray}
where the latter comes from the secondary condition of reality
of the metric
${\rm Im} \{ \partial_t(\tilde{E}^i_a \tilde{E}^{ja} ) \} = 0$
\cite{AshtekarRomanoTate}, and
we assume $\det\tilde{E}>0$ (see \cite{ys-con}).

For later convenience, we also prepare
stronger reality conditions, {\it triad reality} conditions.
The primary and secondary conditions are written respectively as
\begin{eqnarray}
U^i_a := {\rm Im} (\tilde{E}^i_a ) &=& 0,
\label{s-reality1} \\
{\rm and} \qquad
{\rm Im}  ( \partial_t {\tilde{E}^i_a} ) &=& 0.
\label{s-reality2}
\end{eqnarray}
\noindent
Using the equations of motion of $\tilde{E}^i_{a}$,
the gauge constraint (\ref{c-g}),
the metric reality conditions
(\ref{w-reality1}), (\ref{w-reality2-final})
and the primary condition (\ref{s-reality1}),
we see  that  (\ref{s-reality2}) is equivalent to \cite{ys-con}
\begin{eqnarray}
{\rm Re}({\cal A}^a_{0}) &=&
\partial_i(
\null \! \mathop {\vphantom {N}\smash N}\limits ^{}_{^\sim}\!\null
)\tilde{E}^{ia}
+(1 /2e) E^b_i
\null \! \mathop {\vphantom {N}\smash N}\limits ^{}_{^\sim}\!\null
\tilde{E}^{ja} \partial_j\tilde{E}^i_b
+N^{i} {\rm Re}({\cal A}^a_i), \label{s-reality2-final}
\end{eqnarray}
or with un-densitized variables,
\begin{equation}
{\rm Re}({\cal A}^a_{0})=
\partial_i( N)
{E}^{ia}
+N^{i} \, {\rm Re}({\cal A}^a_i).
\label{s-reality2-final2}
\end{equation}
{}From this expression we see that
the secondary triad reality condition
restricts the three components of the ``triad lapse" vector
${\cal A}^a_{0}$.
Therefore (\ref{s-reality2-final}) is
not a restriction on the dynamical variables
($\tilde{E}^i_a $ and ${\cal A}^a_i$)
but on the slicing, which we should impose on each hypersurface.

Throughout the discussion in this article,
we assume that the initial data of
$(\tilde{E}^i_a, {\cal A}^a_i)$ for evolution are solved so as
to satisfy all three constraint equations and the metric
reality condition (\ref{w-reality1}) and (\ref{w-reality2-final}).
Practically, this is
obtained, for
example, by solving ADM constraints and by transforming a
set of initial data to Ashtekar's notation.

\section{Experiments using the Maxwell equation} \label{appB}
In this appendix, using the Maxwell equation of the vacuum field,
we show that the symmetric hyperbolic system does not change
the stability feature drastically.
The result here supports the discussion in \S \ref{Results2}.
More detail analysis can be found in our Paper II \cite{Paper2}.

The Maxwell equation has two constraint equations,
\begin{eqnarray}
C_E&:=&
\partial_i E^i\approx 0,
\quad
C_B:=
\partial_i B^i\approx 0,
\label{MaxCon}
\end{eqnarray}
and two dynamical equations
\begin{eqnarray}
\partial_t E_i&=&
c\epsilon_i{}^{jk} \partial_j B_k
,\quad
\partial_t B_i=
-c\epsilon_i{}^{jk} \partial_j E_k
\label{MaxEQM}
\end{eqnarray}
for the field $(E_i, B_i)$.

Suppose we have adjusted (\ref{MaxEQM}) using the
constraint terms, (\ref{MaxCon}), with a multiplier, $\kappa$.
\begin{eqnarray}
\partial_t E_i&=&
c\epsilon_i{}^{jk} \partial_j B_k
+\kappa_i C_E
,\
\partial_t B_i=
-c\epsilon_i{}^{jk} \partial_j E_k
+\kappa_i C_B
\label{MaxEQM2}
\end{eqnarray}
where $\kappa_i=(\kappa,\kappa,\kappa)$ for simplicity.
This matrix expression
\begin{eqnarray}
\partial_t
\left(\matrix{E_i \cr B_i}\right)
&\cong&
\left(
\matrix{
\delta^{jl}\kappa_i
   &
-c\epsilon_i{}^{jl}
\cr
c\epsilon_i{}^{jl}
&
\delta^{jl}\kappa_i
}
\right)
\partial_{l}
\left(\matrix{E_j \cr B_j}\right)
\label{matri}
\end{eqnarray}
immediately tells us its hyperbolicity depending on $\kappa$ as follows:
The system, (\ref{matri}),  becomes  symmetric hyperbolic form when
$\kappa=0$ (that is the original Maxwell equation),
becomes weakly hyperbolic form when $\kappa = \pm c$, and
becomes strongly hyperbolic otherwise.  The eigenvalues of the
dynamical equation can be written as
$(c,c,-c,-c,\kappa,\kappa)$.

We made a numerical code to demonstrate a propagation of plane
electro-magnetic wave,
\begin{eqnarray}
E^i(x,t)&=&
(0,0,-{1\over \sqrt{2}}\sin({x+y \over \sqrt{2} }-ct)),
\\
B^i(x,t)&=&
(
-{1\over 2}\sin({x+y \over \sqrt{2} }-ct),
{1\over 2}\sin({x+y \over \sqrt{2} }-ct)
,0) \label{maxwellwave}
\end{eqnarray}
in 2-dimensional spacetime with periodic boundary condition.
We use (\ref{maxwellwave}) as our initial data, and monitor
its numerical error during its evolution by evaluating constraint
equations and by checking error from the exact solution.
The error itself is quite small, but as we show in Fig.\ref{figB}
we found the difference due to the multiplier of the adjusted terms $\kappa$.
We see that
the symmetric hyperbolic equation show the best performance for
the stability, but not showing so much different performance from the strongly
hyperbolic system.

\section{Why adjusted equations have better performance?} \label{appC}
Here, we try to explain briefly why the adjusted equations
[(\ref{eqE3}) and (\ref{eqA3}) for Ashtekar's system,
(\ref{MaxEQM2}) for Maxwell equations] reduce the violation
of constraints in the evolution.
The detail explanations and numerical experiments are
in our paper II \cite{Paper2}, and this appendix describes
the essential idea of
the mechanism.

Suppose we have constraint equations, ${\cal C}_1 \approx 0, \,
{\cal C}_2 \approx 0, \cdots$,  in a system.
We, normally,  monitor the error of the evolution by evaluating these
constraint equations on the each constant-time hypersurface.
Such monitoring, on the other hand, can be performed also by
checking the evolution equations of the constraint, which we denote
constraint propagation equations (cf. \cite{Fri-con}).
We, therefore, consider constraint propagation equation of which we 
transformed in Fourier components, $\hat{\cal C}$, 
\begin{equation}
\partial_t \left( \matrix{\hat{\cal C}_1 \cr \hat{\cal C}_2 \cr \vdots }\right)
= M
\left( \matrix{\hat{\cal C}_1 \cr \hat{\cal C}_2 \cr \vdots}\right).
\label{c1}
\end{equation}

The idea here is to estimate the eigenvalues of the
 matrix, $M$, after we took 
its leading order quantitiy 
in linealization against a particular background.  
Clearly, if all the real part of the eigenvalues are negative,
then all constraints decays to zero along to the system's evolution.
In our paper II \cite{Paper2}, we show that such a case can be obtained
by adding `adjusted terms' both for Ashtekar's and Maxwell's systems.
There we also show examples of unstable evolution by choosing adjusted
terms which produce positive eigenvalues of $M$.  The imaginary part of
the eigenvalues are also supposed to contribute the appearance of the
phase differences of the system.

In this point, we can say that adjusted terms are responsible for
obtaining the stable and/or accurate evolution system, and this is a
way to control the stability of simulation, which effects more than the
system's hyperbolicity.


\if\answ\nofig
\begin{figure}[h]
\fi
\if\answ\onecol
\begin{figure}[thb]
\setlength{\unitlength}{1in}
\begin{picture}(7.0,6.50)
\put(0.50,3.00){\epsfxsize=2.8in \epsfysize=1.6in
\epsffile{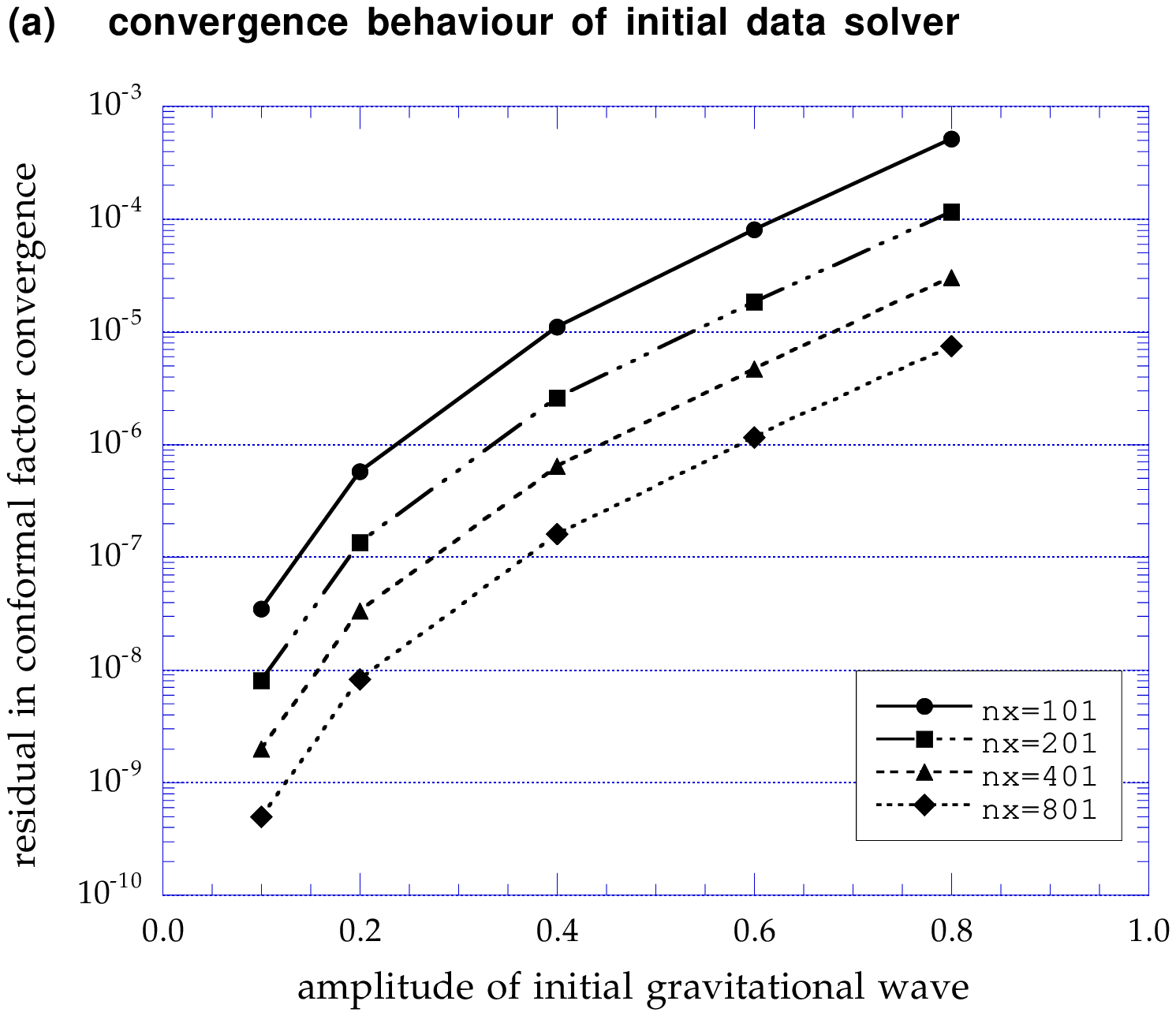}}
\put(0.25,0.25){\epsfxsize=3.0in \epsfysize=1.8in \epsffile{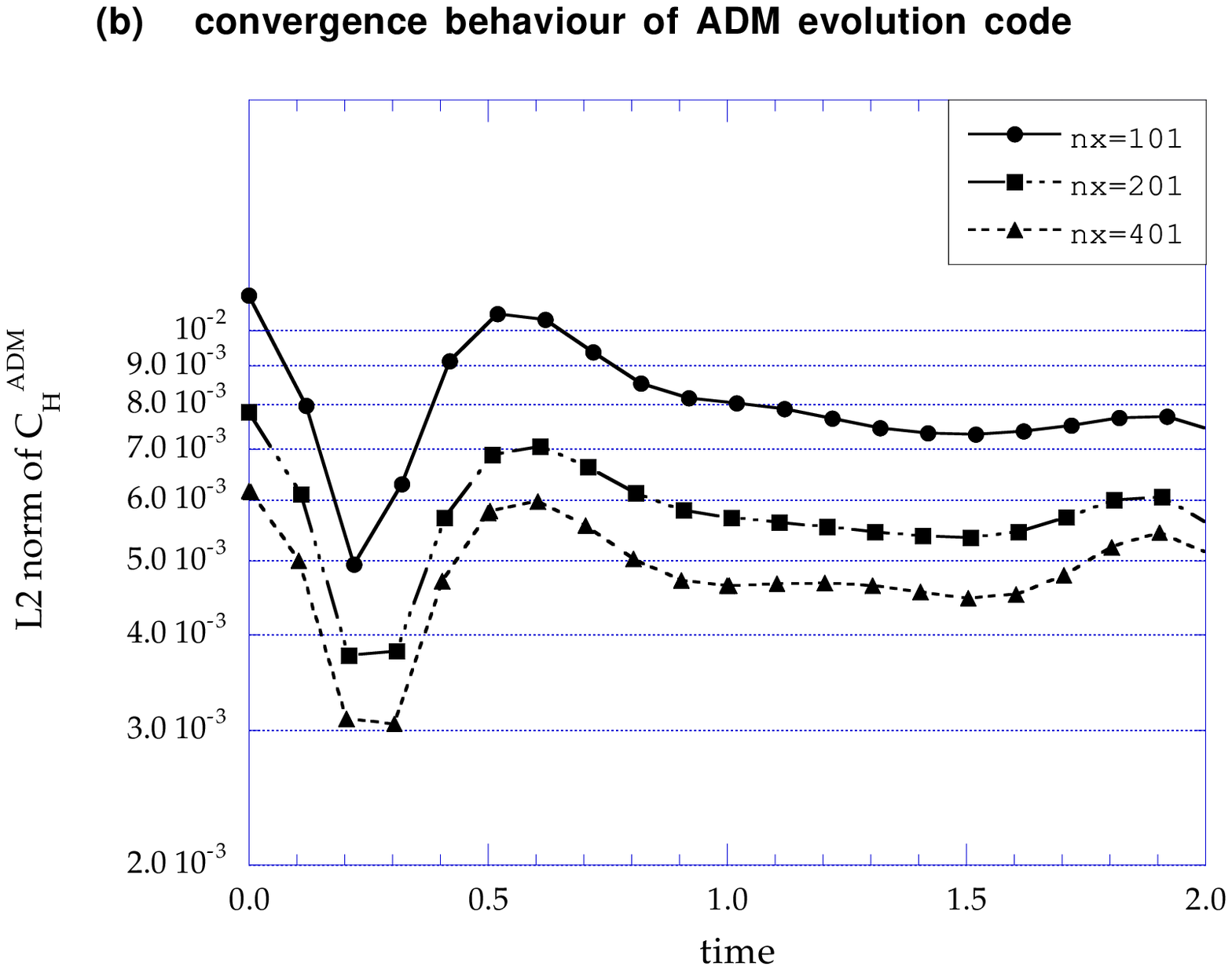}}
\put(3.50,0.25){\epsfxsize=3.0in \epsfysize=1.8in \epsffile{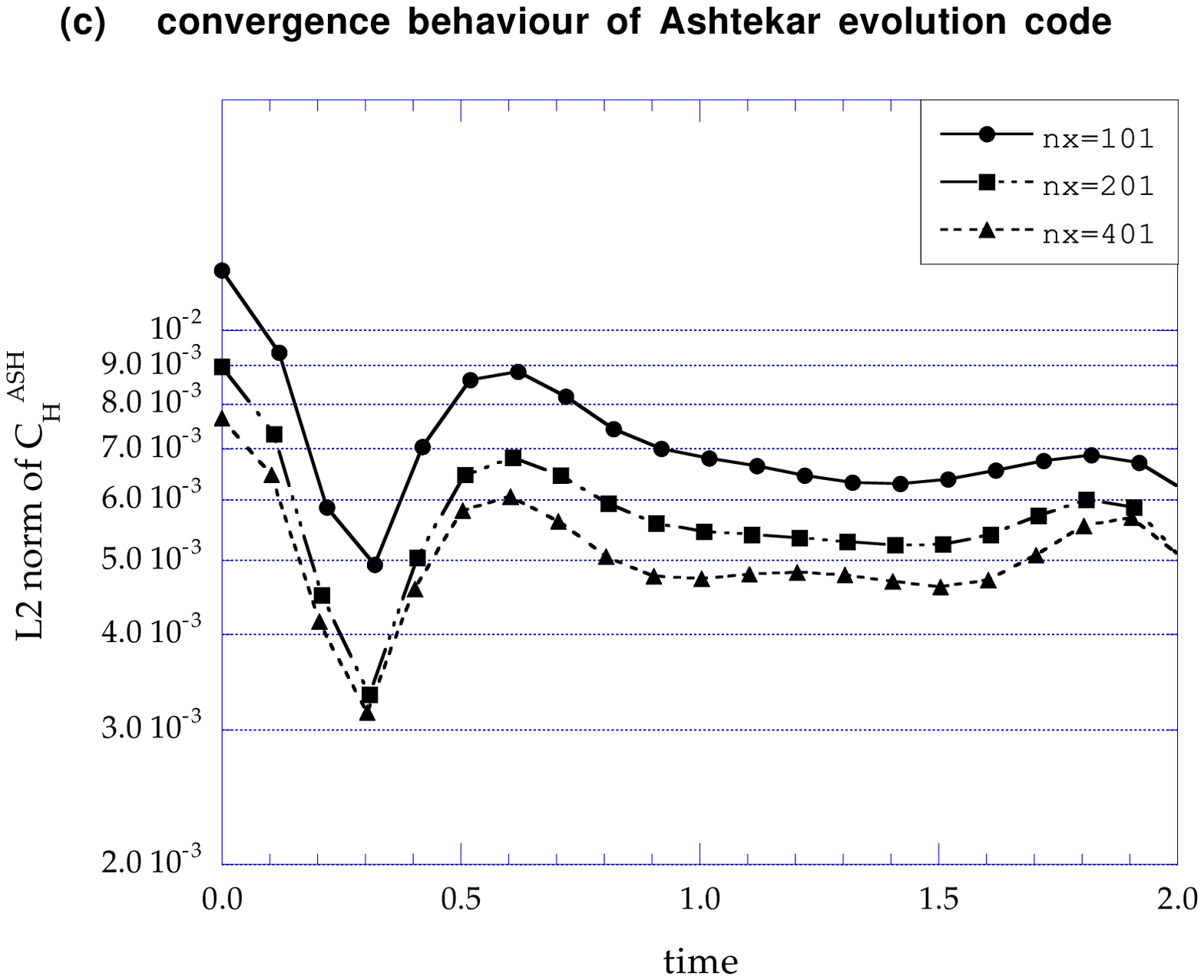}}
\end{picture}
\fi
\if\answ\prepri
\begin{figure}[p]
\setlength{\unitlength}{1in}
\begin{picture}(7.0,5.50)
\put(0.25,3.25){\epsfxsize=3.0in \epsfysize=2.15in
\epsffile{fig1a_initconv.eps} }
\put(0.25,0.25){\epsfxsize=3.0in \epsfysize=2.15in
\epsffile{fig1b_admconv.eps}}
\put(3.50,0.25){\epsfxsize=3.0in \epsfysize=2.15in
\epsffile{fig1c_ashconv.eps}}
\end{picture}
\fi

\caption[conv_test]{Examples of the convergence tests.
(a) Convergence of the initial data solver [Hamiltonian constraint
(\ref{bchamilt_vac}) solver].  We plot the residual of the conformal factor,
L2 norm of
$\left| (\psi_n - \psi_{n-1})/\psi_n \right|^2$, when it converges,
where $n$ is the iteration number in ICCG routine.
The horizontal value is the amplitude of the gravitational wave
[$a$ in eq.(\ref{plusmetric})] where we assume
$+$-mode single pulse wave
and fix $b=2.0, c=0.0$ in
eq.(\ref{plusmetric}), and $K_0=-0.025$.
According to the resolutions (grid points = 101, $\cdots$, 801
   for the range of $x=[-5,+5]$), we see
second order convergence.
(b) Convergence behavior of ADM evolution code.
L2 norms of ${\cal C}^{\rm ADM}_H$ is plotted for the above initial data
for the amplitude $a=0.2$. We applied geodesic slicing condition ($N=1$).
(c) Convergence behavior of Ashtekar evolution code.
L2 norms of ${\cal C}^{\rm ASH}_H$ is plotted for the above initial data
for the amplitude $a=0.2$. We applied geodesic slicing condition ($N=1$).
We can see clearly that the error norms in evolutions will
decrease in high resolution cases.
}
\label{conv_test}
\end{figure}

\if\answ\nofig
\begin{figure}[h]
\fi
\if\answ\onecol
\begin{figure}[phtb]
\setlength{\unitlength}{1in}
\begin{picture}(8.0,7.0)
\put(0.10,3.50) {\epsfxsize=3.0in \epsfysize=1.8in \epsffile{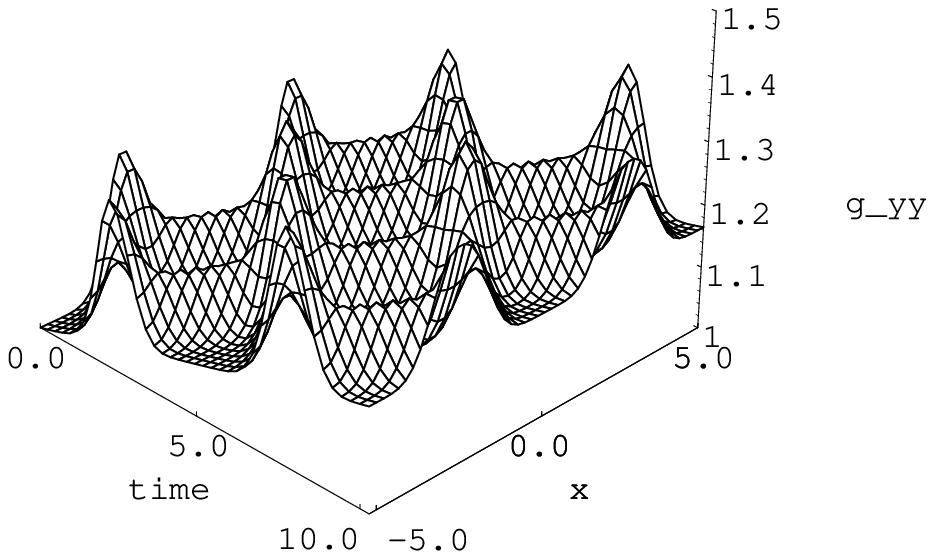} }
\put(3.50,3.50) {\epsfxsize=3.0in \epsfysize=1.8in \epsffile{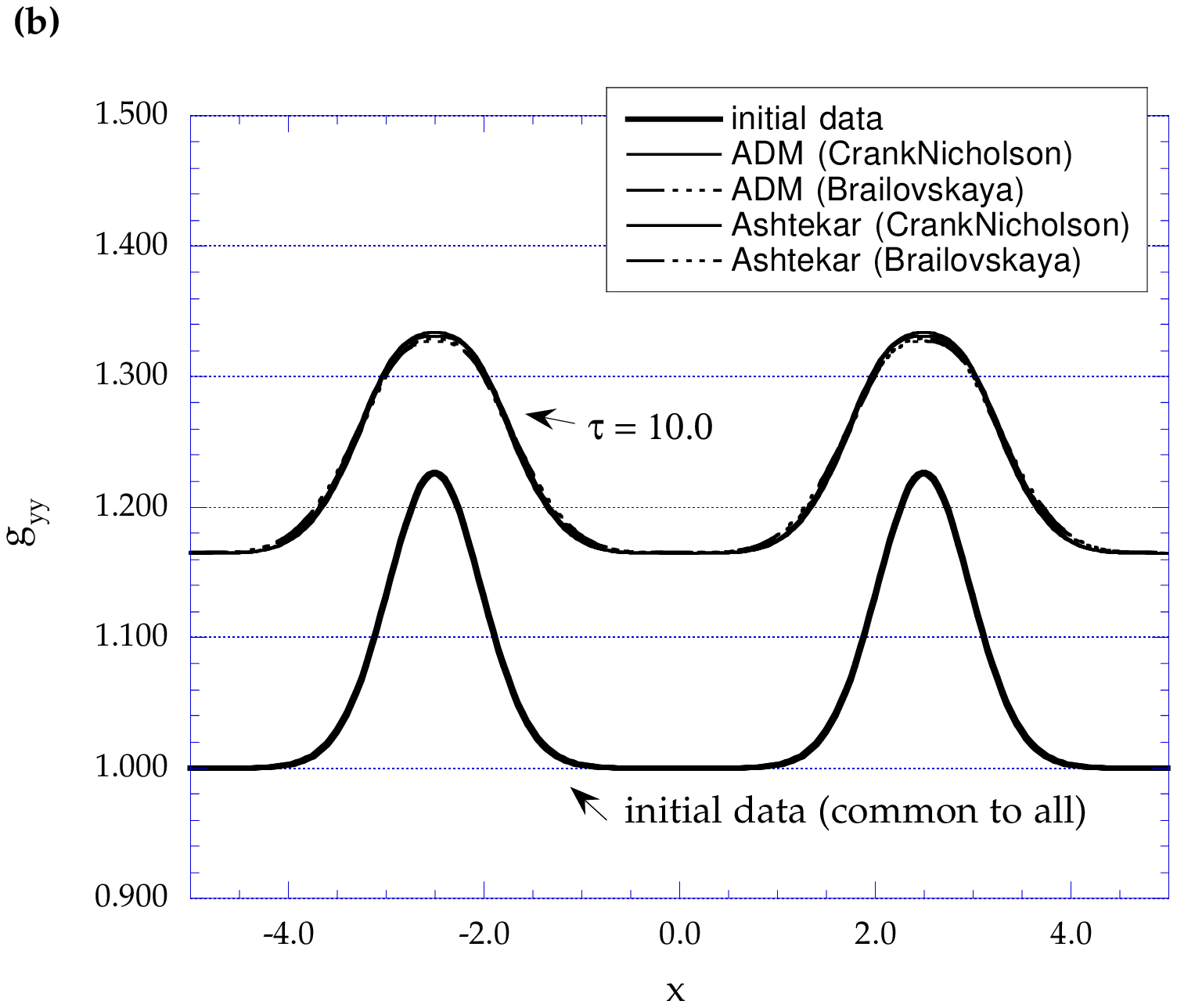} }
\put(0.10,0.25) {\epsfxsize=3.0in \epsfysize=1.8in \epsffile{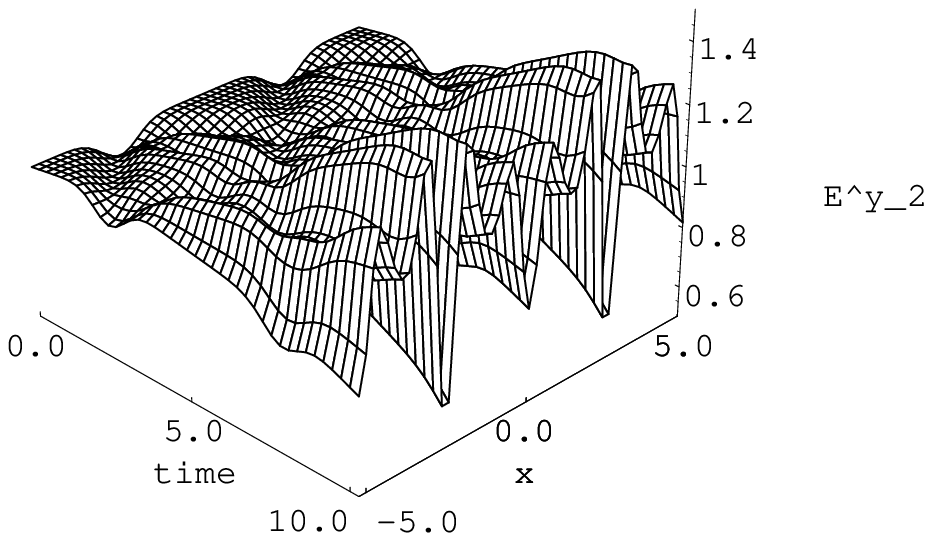} }
\put(3.50,0.25) {\epsfxsize=3.0in \epsfysize=1.8in \epsffile{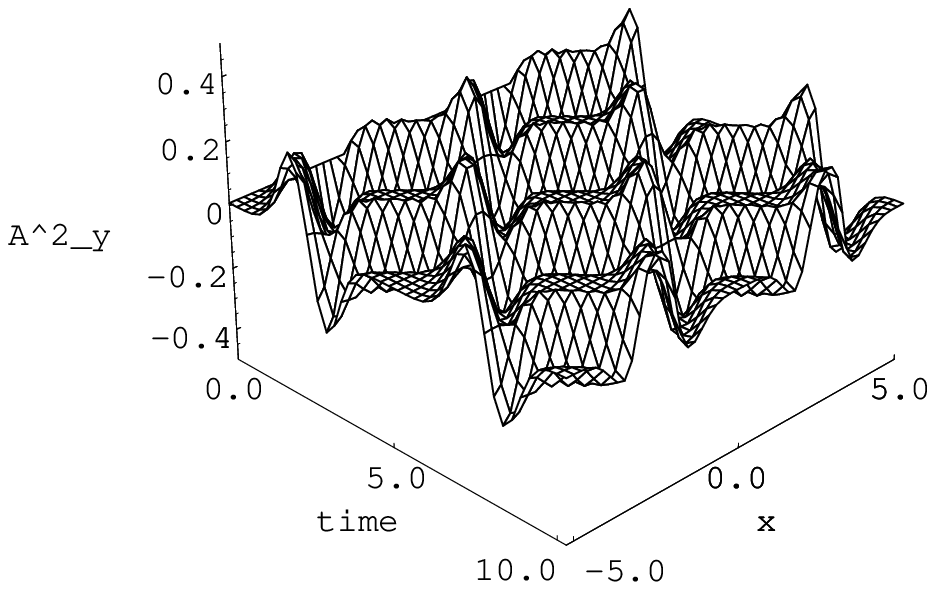} }
\end{picture}
\fi
\if\answ\prepri
\begin{figure}[p]
\setlength{\unitlength}{1in}
\begin{picture}(8.0,7.0)
\put(0.10,3.50) {\epsfxsize=3.0in \epsfysize=1.8in \epsffile{fig2a_gyy.eps} }
\put(3.50,3.50) {\epsfxsize=3.0in \epsfysize=1.8in \epsffile{fig2b.eps} }
\put(0.10,0.25) {\epsfxsize=3.0in \epsfysize=1.8in \epsffile{fig2c_E22.eps} }
\put(3.50,0.25) {\epsfxsize=3.0in \epsfysize=1.8in \epsffile{fig2d_A22.eps} }
\end{picture}
\fi

\caption[conv_test]{
Images of gravitational wave propagation and comparisons of dynamical
behaviour of Ashtekar's variables and ADM variables.  We applied the
same initial data of two $+$-mode pulse waves ($a=0.2, b=2.0, c=\pm2.5$ in
eq.(\ref{plusmetric}) and $K_0=-0.025$), and the same slicing
condition, the standard geodesic slicing condition ($N=1$).
Fig.(a) [top left]  is the image of three-metric component
$g_{yy}$ of a function of proper time $\tau$ and coordinate $x$.
This behaviour can be seen identically both in ADM and Ashtekar evolutions,
and both with Brailovskaya and Crank-Nicholson  time integrating scheme.
Fig.(b) explains this fact by comparing the snapshot of $g_{yy}$
at the same propertime slice ($\tau=10$), where four lines at $\tau=10$
are looked identically.
Figs.(c) and (d) [botoms]
are of the real part of the densitized triad $\tilde{E}^y_2$, and the
real part of the
connection ${\cal A}^2_y$, respectively, obtained from Ashtekar variables'
evolution.
}
\label{admash}
\end{figure}

\if\answ\nofig
\begin{figure}[h]
\fi
\if\answ\onecol
\begin{figure}[h]
\setlength{\unitlength}{1in}
\begin{picture}(7.0,3.0)
\put(0.25,0.25){\epsfxsize=3.0in \epsfysize=1.8in \epsffile{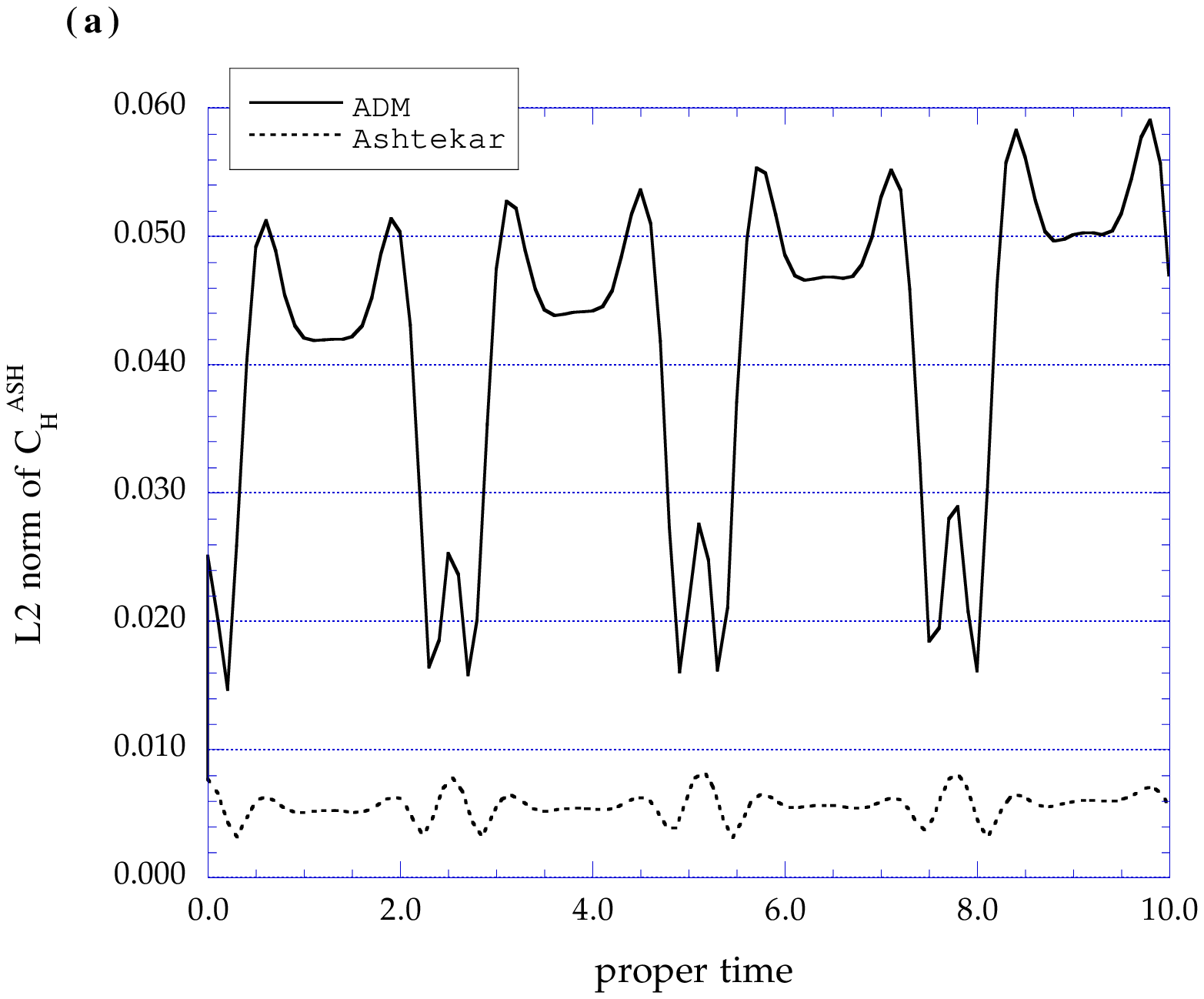} }
\put(3.75,0.25){\epsfxsize=3.0in \epsfysize=1.8in \epsffile{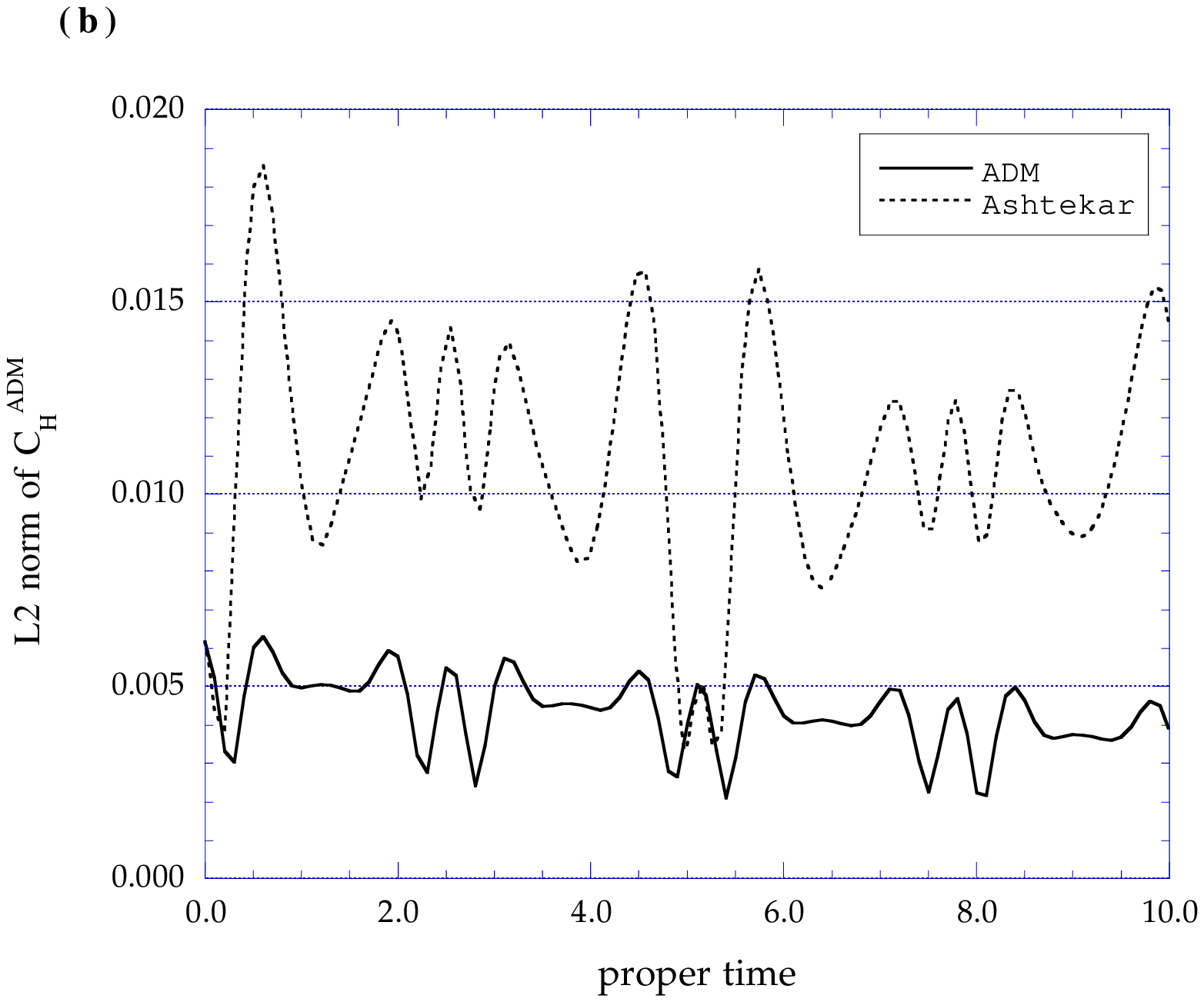} }
\end{picture}
\fi
\if\answ\prepri
\begin{figure}[p]
\setlength{\unitlength}{1in}
\begin{picture}(7.0,7.0)
\put(1.5,4.25){\epsfxsize=4.0in \epsfysize=2.38in \epsffile{fig3a.eps} }
\put(1.5,0.25){\epsfxsize=4.0in \epsfysize=2.38in \epsffile{fig3b.eps} }
\end{picture}
\fi

\caption[conv_test]{
Comparisons of the constraint violation by Ashtekar's equation with that of 
ADM.
(a) L2 norm of the Ashtekar's Hamiltonian constraint equation,
${\cal C}_H^{\rm ASH}$ as a function of averaged proper time.
(b) L2 norm of the ADM's Hamiltonian constraint equation, ${\cal C}_H^{\rm 
ADM}$
as a function of averaged proper time.
The plots are of the same parameters with those of Fig.\ref{admash}.
}
\label{admash2}
\end{figure}

\if\answ\nofig
\begin{figure}[h]
\fi
\if\answ\onecol
\begin{figure}[phtb]
\setlength{\unitlength}{1in}
\begin{picture}(7.0,6.0)
\put(.25,3.25) {\epsfxsize=3.0in \epsfysize=2.0in \epsffile{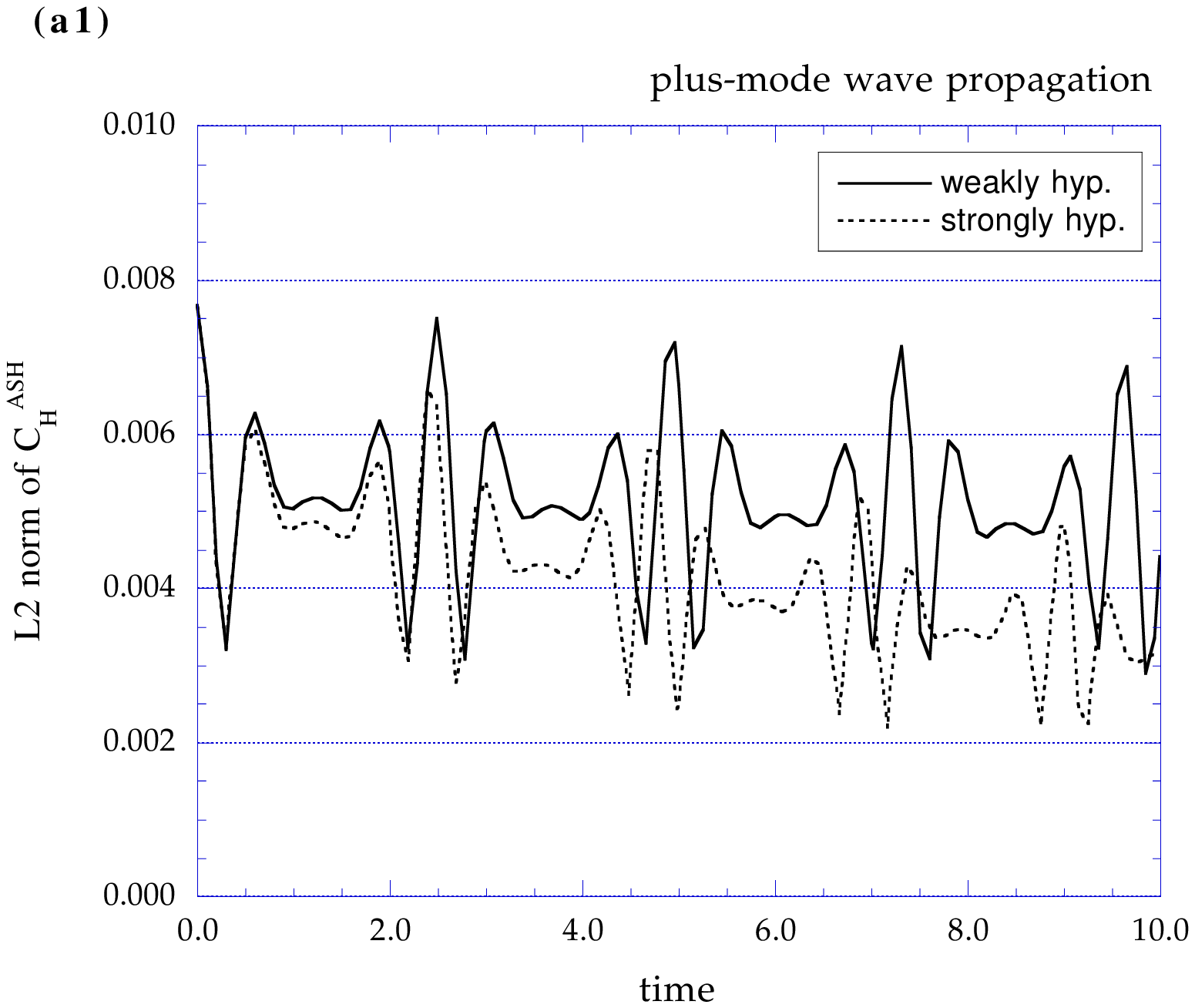} }
\put(3.5,3.25) {\epsfxsize=3.0in \epsfysize=2.0in \epsffile{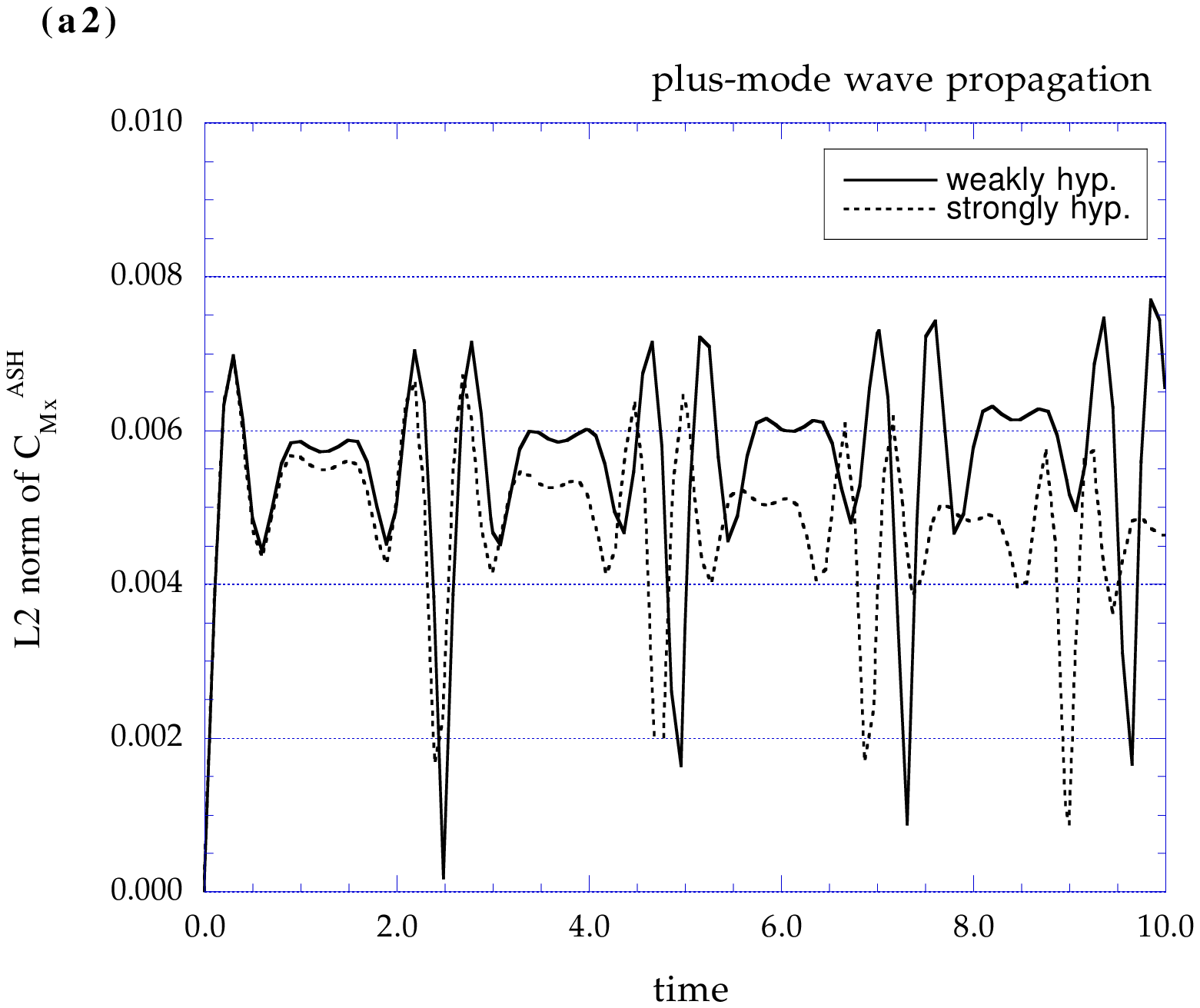} }
\put(0.25,0.25){\epsfxsize=3.0in \epsfysize=2.0in \epsffile{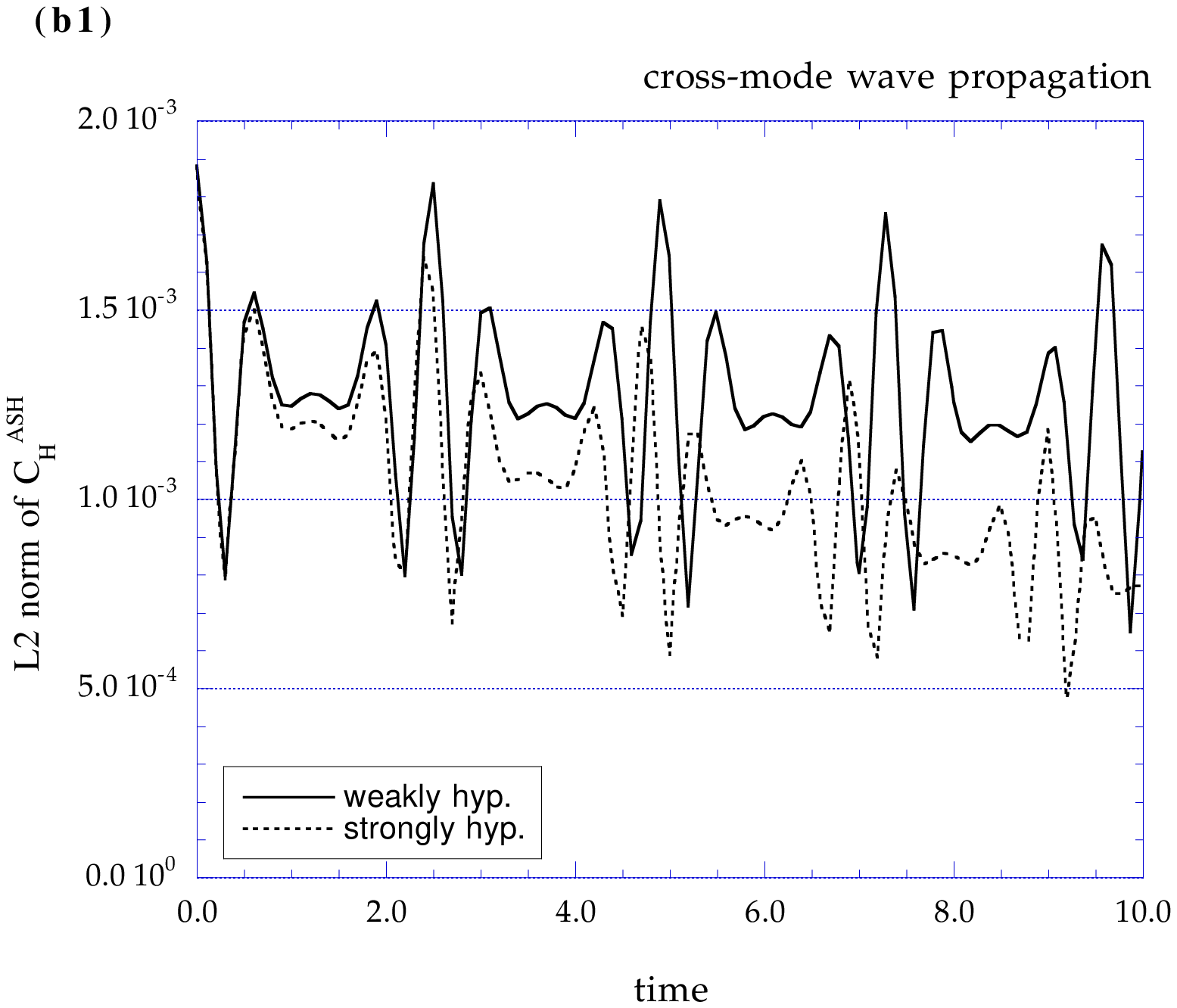} }
\put(3.5,0.25) {\epsfxsize=3.0in \epsfysize=2.0in \epsffile{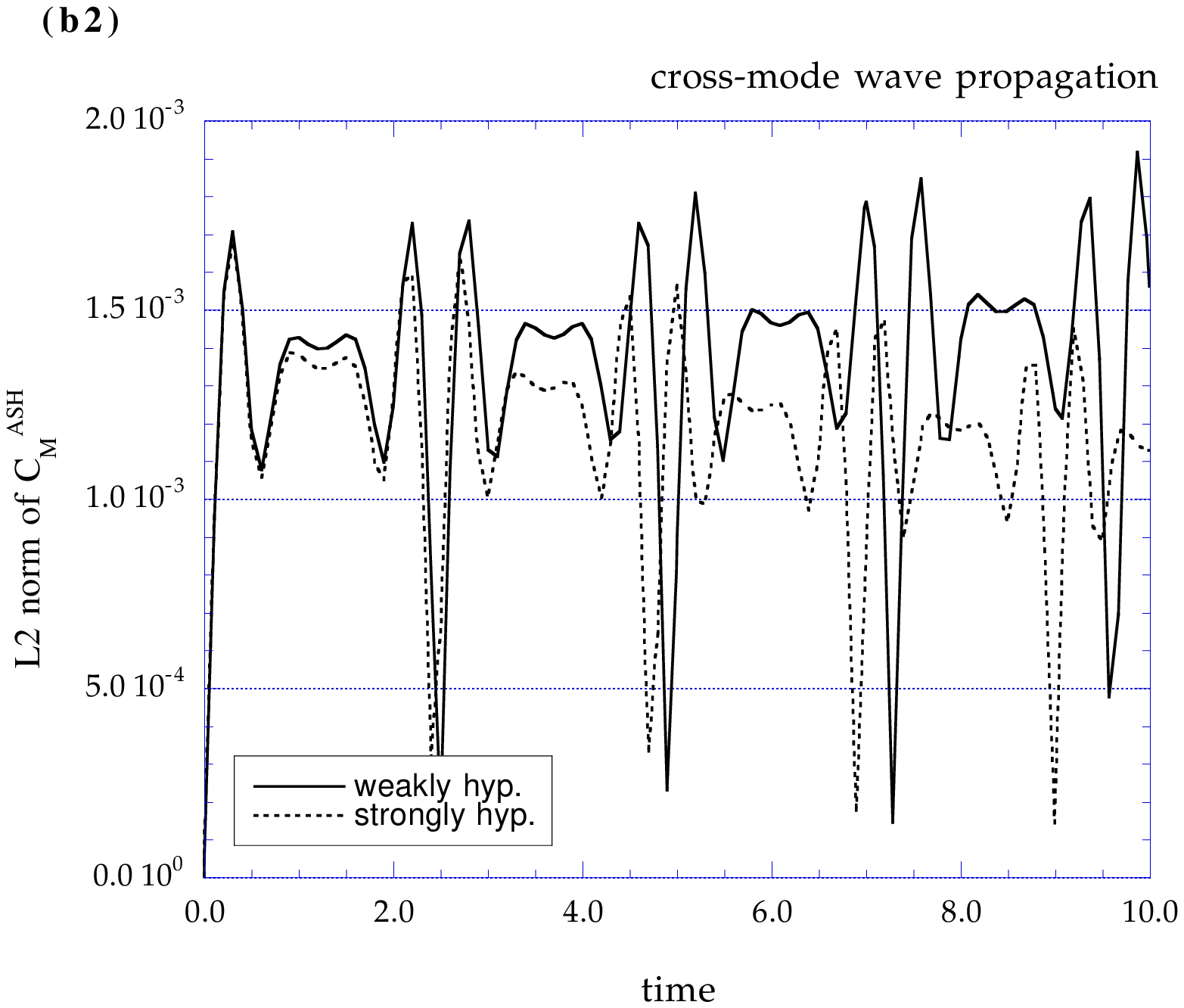} }
\end{picture}
\fi
\if\answ\prepri
\begin{figure}[p]
\setlength{\unitlength}{1in}
\begin{picture}(7.0,6.0)
\put(.25,3.25) {\epsfxsize=3.0in \epsfysize=2.0in \epsffile{fig4a1.eps} }
\put(3.5,3.25) {\epsfxsize=3.0in \epsfysize=2.0in \epsffile{fig4a2.eps} }
\put(0.25,0.25){\epsfxsize=3.0in \epsfysize=2.0in \epsffile{fig4b1.eps} }
\put(3.5,0.25) {\epsfxsize=3.0in \epsfysize=2.0in \epsffile{fig4b2.eps} }
\end{picture}
\fi

\caption[conv_test]{
Comparisons of the strongly hyperbolic system (Ashtekar II)
with the weakly hyperbolic system (Ashtekar original)
($\null \! \mathop {\vphantom {N}\smash N}\limits ^{}_{^\sim}\!\null =1$
slice).
Figs. (a)s are of  $+$-mode waves
($a=0.2, b=2.0, c=\pm2.5$ in
eq.(\ref{plusmetric}),
while (b)s are of $\times$-mode
waves ($a=0.1, b=2.0, c=\pm2.5$ in
eq.(\ref{crossmetric}), in a background spacetime with
$K_0=0.025$.  We plot the L2 norm of
the Hamiltonian and momentum
constraints, ${\cal C}^{\rm ASH}_H$ and ${\cal C}^{\rm ASH}_M$, for each cases.
We see from all of them
that strongly hyperbolic system improves the violation
of the constraints.
}
\label{hikaku_Ne}
\end{figure}

\if\answ\nofig
\begin{figure}[h]
\fi
\if\answ\onecol
\begin{figure}[phtb]
\setlength{\unitlength}{1in}
\begin{picture}(7.0,6.0)
\put(.25,3.25) {\epsfxsize=3.0in \epsfysize=2.0in \epsffile{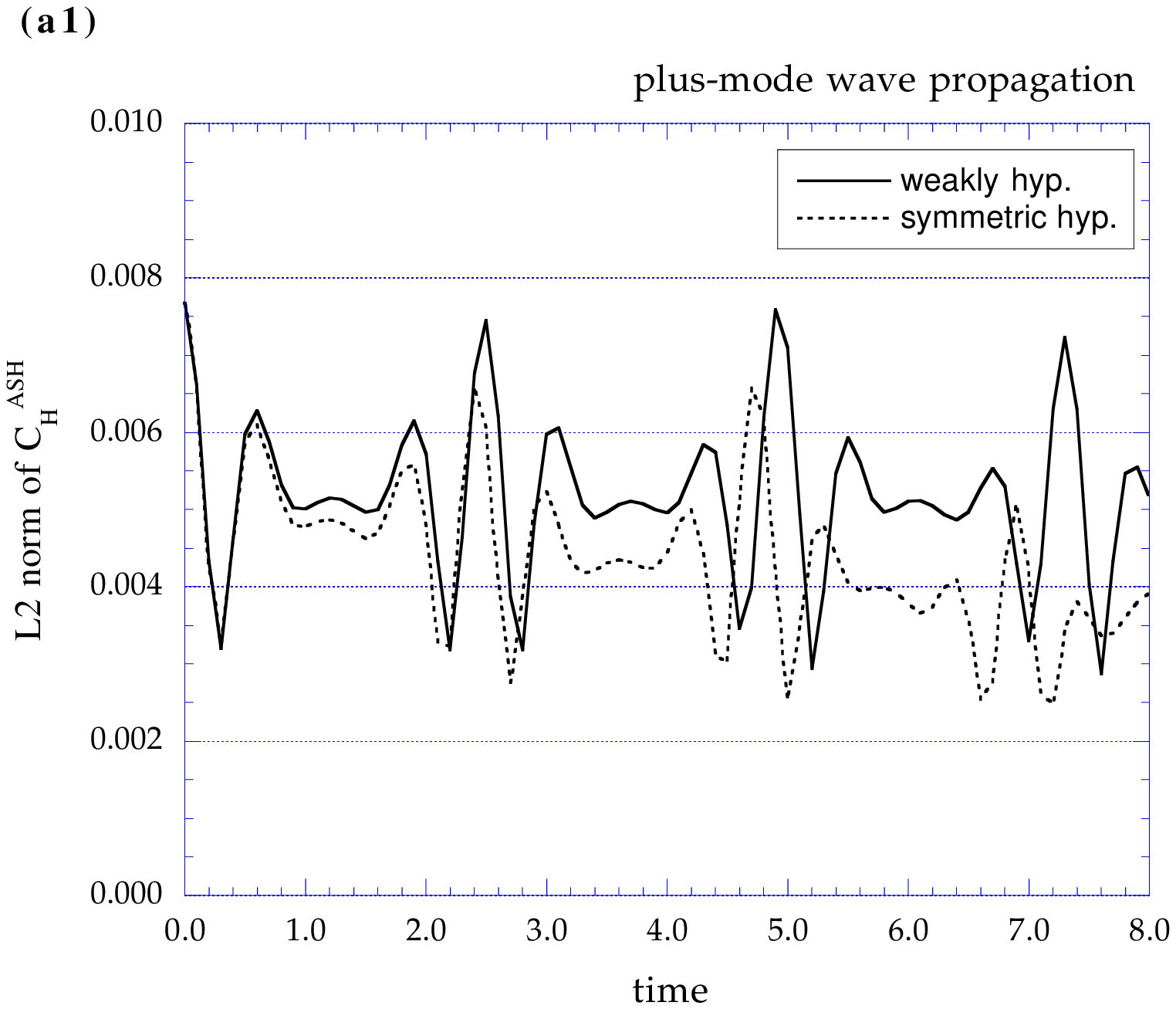} }
\put(3.5,3.25) {\epsfxsize=3.0in \epsfysize=2.0in \epsffile{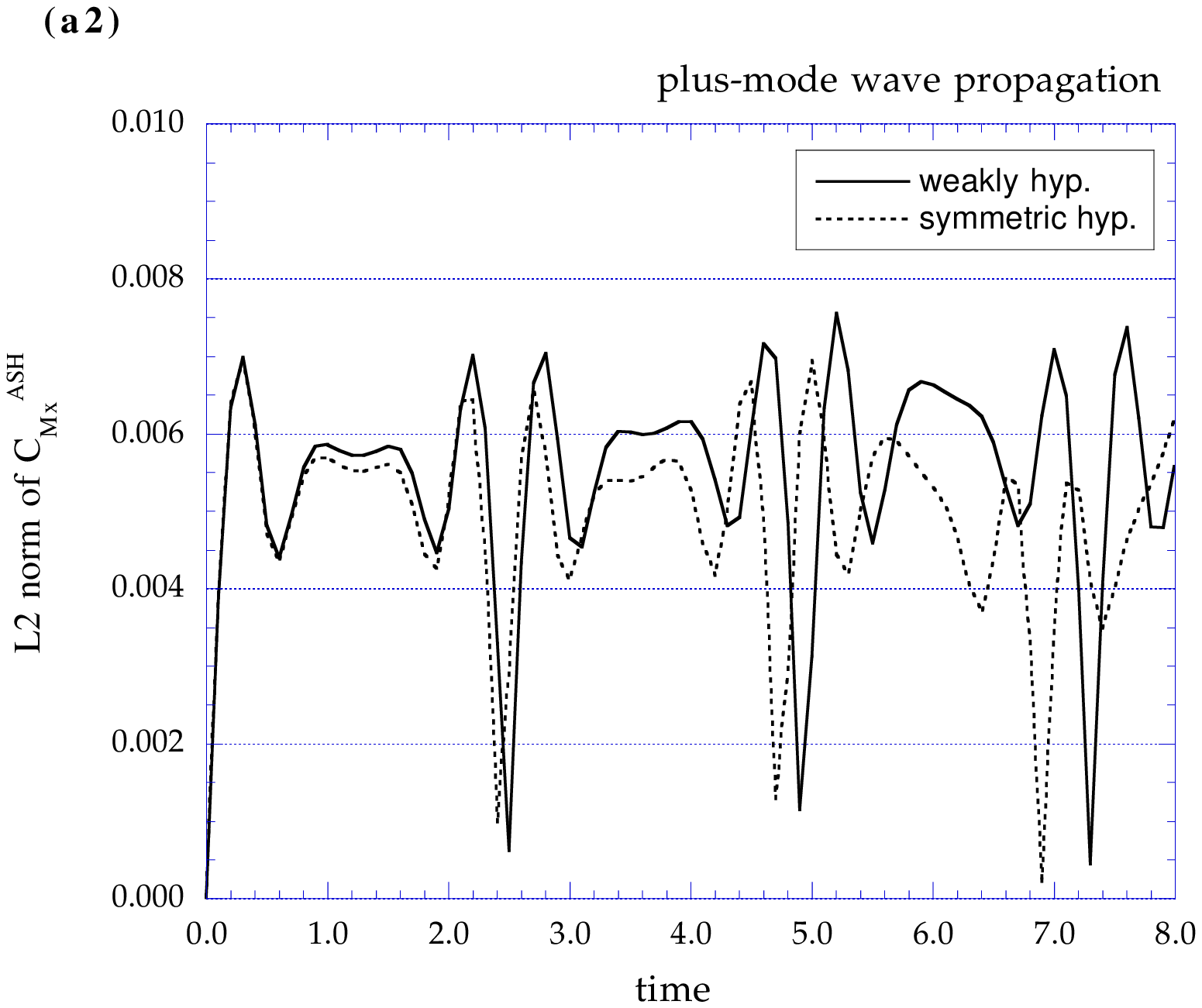} }
\put(0.25,0.25){\epsfxsize=3.0in \epsfysize=2.0in \epsffile{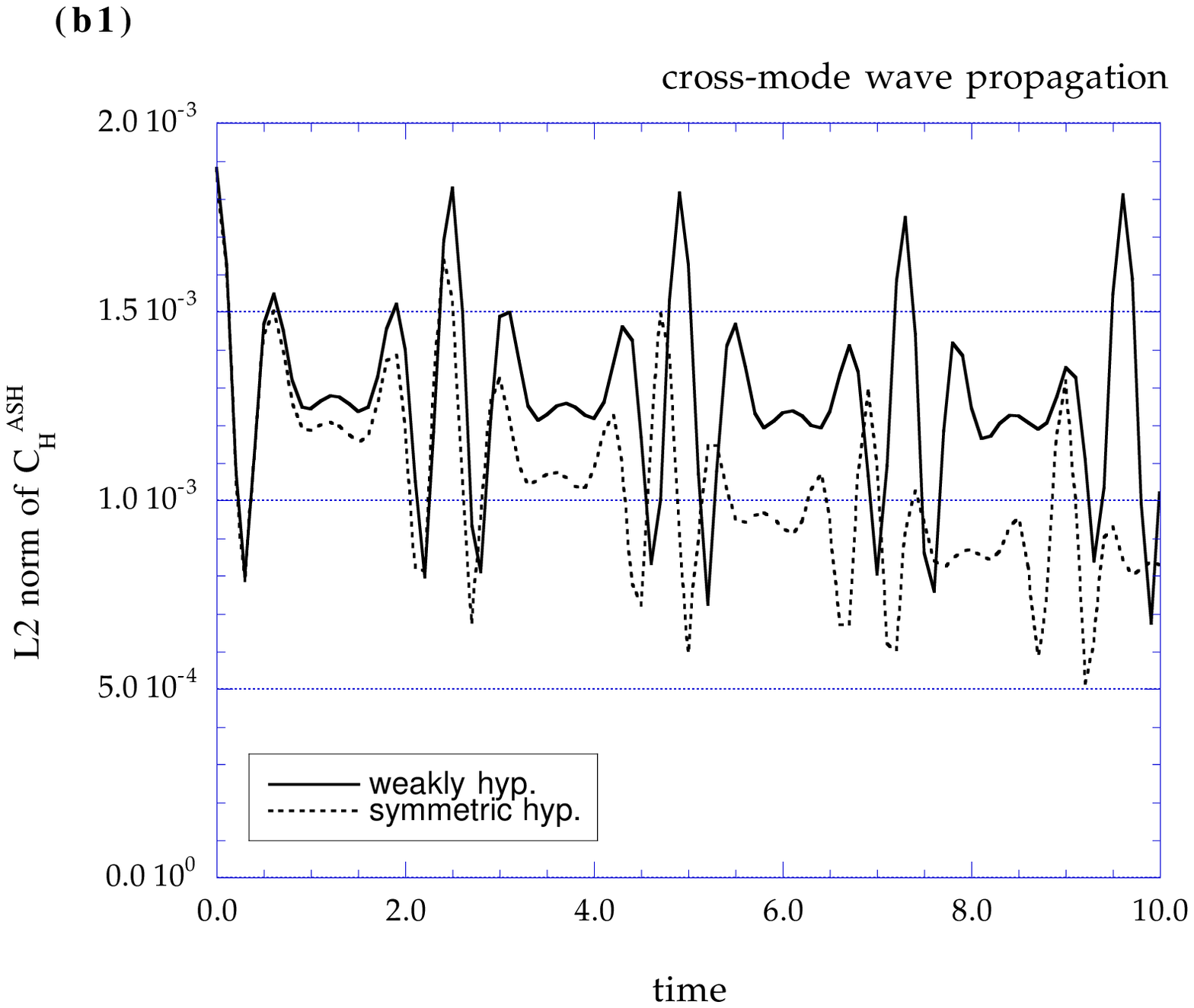} }
\put(3.5,0.25) {\epsfxsize=3.0in \epsfysize=2.0in \epsffile{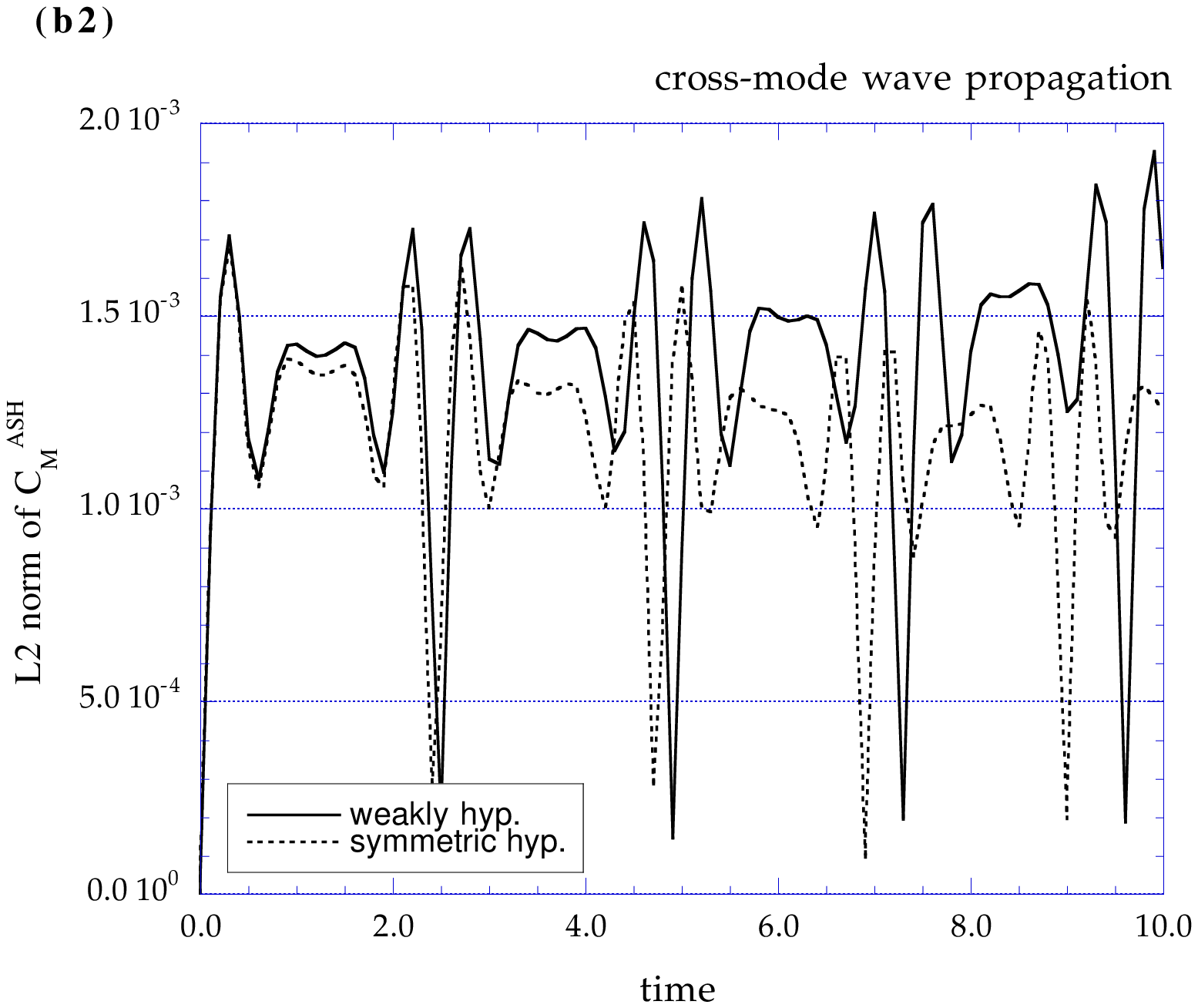} }
\end{picture}
\fi
\if\answ\prepri
\begin{figure}[p]
\setlength{\unitlength}{1in}
\begin{picture}(7.0,6.0)
\put(.25,3.25) {\epsfxsize=3.0in \epsfysize=2.0in \epsffile{fig5a1.eps} }
\put(3.5,3.25) {\epsfxsize=3.0in \epsfysize=2.0in \epsffile{fig5a2.eps} }
\put(0.25,0.25){\epsfxsize=3.0in \epsfysize=2.0in \epsffile{fig5b1.eps} }
\put(3.5,0.25) {\epsfxsize=3.0in \epsfysize=2.0in \epsffile{fig5b2.eps} }
\end{picture}
\fi

\caption[conv_test]{
Comparisons of the symmetric hyperbolic system (Ashtekar III)
with the weakly hyperbolic system (Ashtekar original)
($N=1$ slice), in the same way with Fig.\ref{hikaku_Ne}.
We applied the same parameters with those of Fig.\ref{hikaku_Ne}.
Figs. (a1) and (a2) are of $+$-mode waves propagation and L2 norm of
${\cal C}^{\rm ASH}_H$ and ${\cal C}^{\rm ASH}_M$, respectively.
Figs. (b1) and (b2) are of $\times$-mode waves propagation
and L2 norm of
${\cal C}^{\rm ASH}_H$ and ${\cal C}^{\rm ASH}_M$,
respectively.
We see from all of them
that symmetric hyperbolic system improves the violation of the
constraints.
}
\label{hikaku_N1}
\end{figure}

\if\answ\nofig
\begin{figure}[h]
\fi
\if\answ\onecol
\begin{figure}[h]
\setlength{\unitlength}{1in}
\begin{picture}(7.0,3.0)
\put(0.25,0.25){\epsfxsize=3.0in \epsfysize=1.8in \epsffile{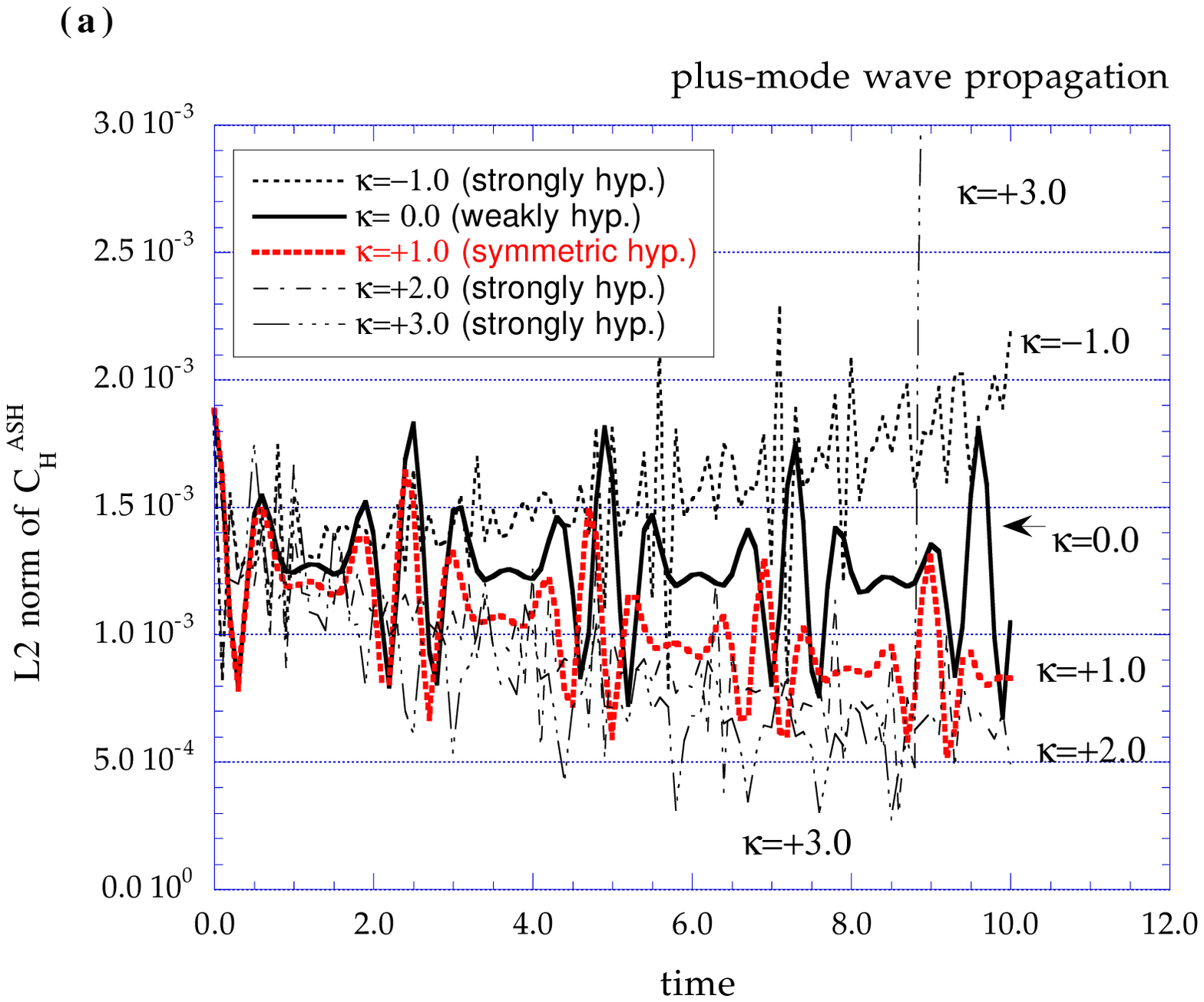} }
\put(3.75,0.25){\epsfxsize=3.0in \epsfysize=1.8in \epsffile{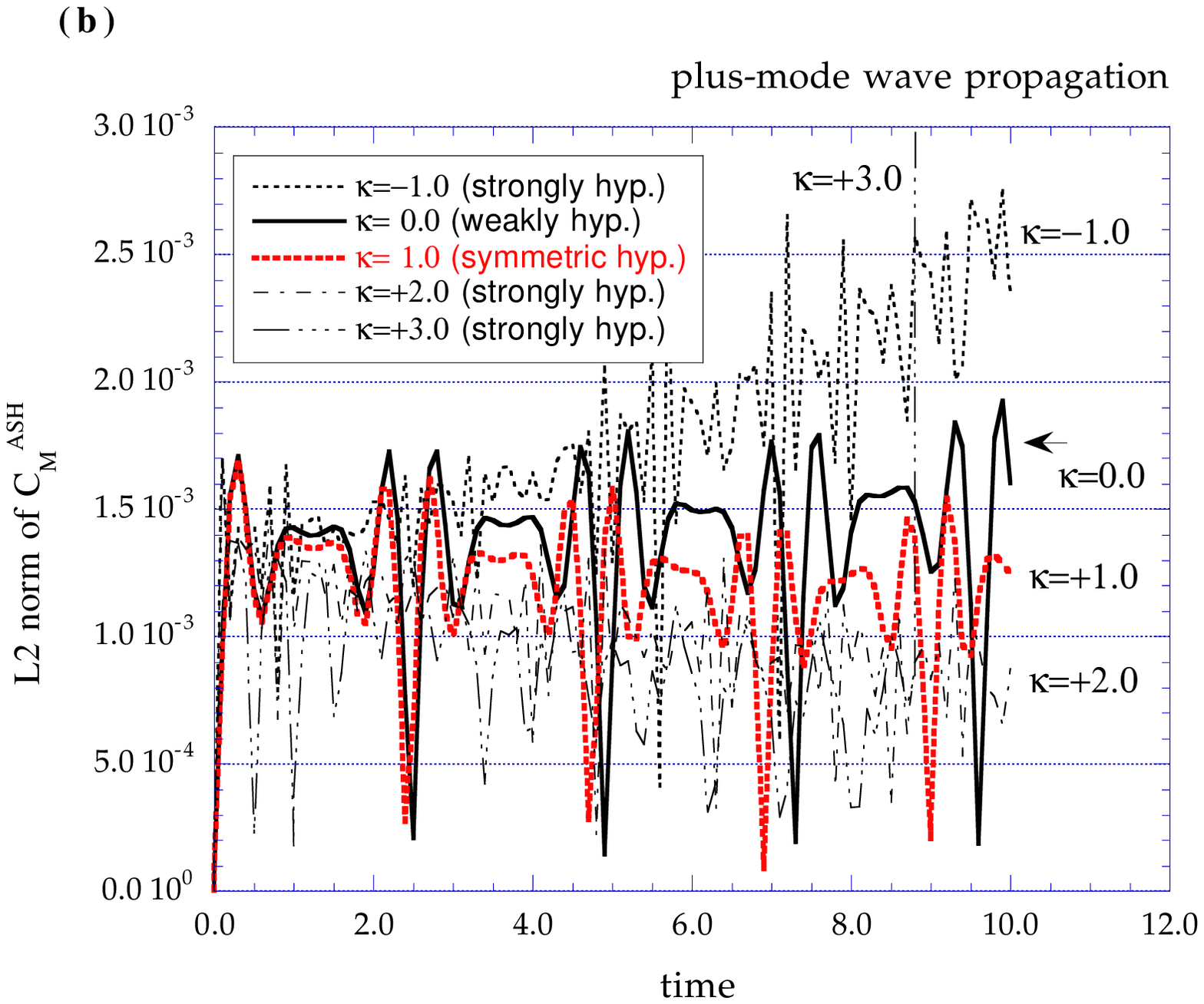} }
\end{picture}
\fi
\if\answ\prepri
\begin{figure}[p]
\setlength{\unitlength}{1in}
\begin{picture}(7.0,6.0)
\put(1.5,3.25){\epsfxsize=4.0in \epsfysize=2.38in \epsffile{fig6a.eps} }
\put(1.5,0.25){\epsfxsize=4.0in \epsfysize=2.38in \epsffile{fig6b.eps} }
\end{picture}
\fi

\caption[conv_test]{
Comparisons of `adjusted' system  with the different
multiplier, $\kappa$, in eqs. (\ref{eqE3}) and (\ref{eqA3}).
The model was $+$-mode pulse waves ($a=0.1, b=2.0, c=\pm2.5$ in
eq.(\ref{plusmetric}) in a background $K_0=-0.025$.
Plots are of
L2 norm of the Hamiltonian and momentum constraint equations,
${\cal C}_H^{\rm ASH}$
and ${\cal C}_M^{\rm ASH}$ [Figs.(a) and (b)] respectively.
We see some $\kappa$ produce better performance than the symmetric
hyperbolic system.
}
\label{kappa}
\end{figure}

\if\answ\nofig
\begin{figure}[h]
\fi
\if\answ\onecol
\begin{figure}[h]
\setlength{\unitlength}{1in}
\begin{picture}(7.0,3.0)
\put(0.25,0.25){\epsfxsize=3.0in \epsfysize=1.8in \epsffile{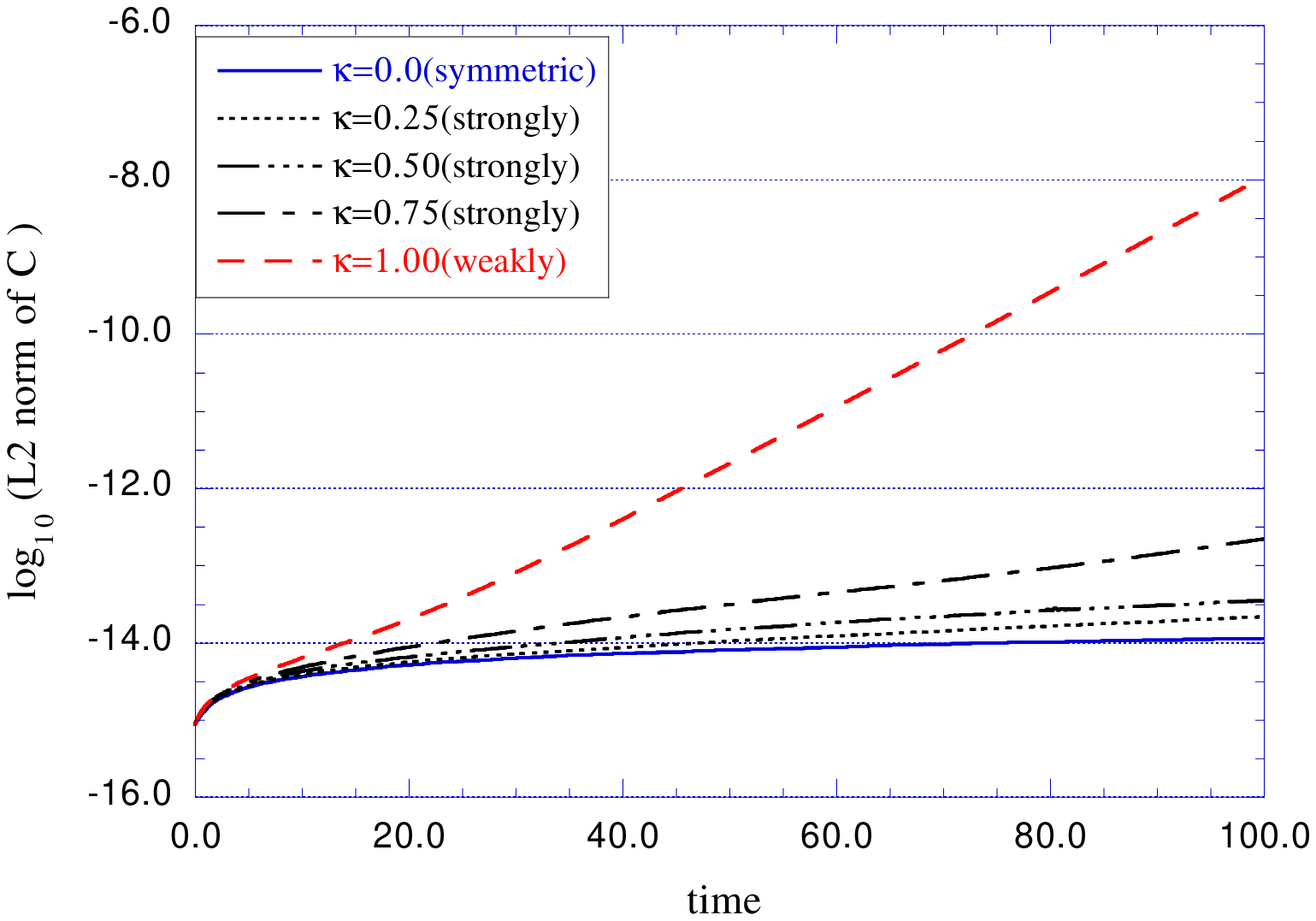} }
\end{picture}
\fi
\if\answ\prepri
\begin{figure}[p]
\setlength{\unitlength}{1in}
\begin{picture}(7.0,3.0)
\put(1.5,0.25){\epsfxsize=4.0in \epsfysize=2.38in \epsffile{figB.eps} }
\end{picture}
\fi

\caption[conv_test]{
Comparisons of `adjusted' system with the different
multiplier, $\kappa$, in eq. (\ref{MaxEQM2}).
Plots are of
L2 norm of the constraint equations, (\ref{MaxCon}).
}
\label{figB}
\end{figure}


\begin{thebibliography}{99}
\bibitem{hyp-comparisons}
Recently, we heard that J. Bardeen and L. Buchman
are preparing a numerical comparison
between Bona-Mass\'o (97) and Anderson-York (99) formulations.
(see also \cite{hern}).
\end{thebibliography}
\end{document}